\documentclass[11pt]{article}

\usepackage{epsfig,latexsym}
\usepackage{amsmath}
\usepackage{graphicx}
\usepackage{lineno}
\usepackage{amsfonts}
\newcommand{\qed}{\hfill $\Box$}

\newtheorem{defi}{Definition}
\newtheorem{theo}{Theorem}
\newtheorem{lemm}{Lemma}
\newtheorem{coro}{Corollary}

\newtheorem{prop}{Proposition}
\newtheorem{obse}{Observation}

\title{The stable set polytope of ($P_6$,triangle)-free graphs and new facet-inducing graphs}

\author{
Raffaele Mosca\thanks{Dipartimento di Economia, Universit\'a degli Studi ``G. D'Annunzio'', Pescara, Italy.
E-mail: r.mosca@unich.it}}

\begin{document}

\maketitle

%\def\inst#1{$^{#1}$}
%%
%\title{The stable set polytope for ($P_6$,triangle)-free graphs (and new facet-inducing graphs)}

%\author{Raffaele Mosca}

%\begin{document}

%\maketitle

%\begin{center}
%{\footnotesize
%Dipartimento di Scienze, Universit\'a degli Studi "G. d'Annunzio", Pescara 65127, Italy.\\
%\texttt{r.mosca@unich.it} }
%\end{center}

\begin{abstract}

%[{\em Ad Laudem Domini.}]

The stable set polytope of a graph $G$, denoted as STAB($G$), is the convex hull of all the incidence vectors of stable sets of $G$. To describe a linear system which defines STAB($G$) seems to be a difficult task in the general case. In this paper we present a complete description of the stable set polytope of ($P_6$,triangle)-free graphs (and more generally of ($P_6$,paw)-free graphs). For that we combine different tools, in the context of a well known result of Chv\'atal \cite{Chvatal1975} which allows to focus just on prime facet-inducing graphs, with particular reference to a structure result on prime ($P_6$,triangle)-free graphs due to Brandst\"adt et al. \cite{BraKleMah2005}. Also we point out some peculiarities of new facet-inducing graphs detected along this study with the help of a software. \\

\end{abstract}

\section{Introduction}

Let $G = (V,E)$ be a graph of $n$ vertices. A {\em stable set} of $G$ is a set of pairwise nonadjacent vertices of $G$. Each
stable set $S$ of $G$ is characterized by its incidence vector,
that is an $n$-vector whose $i$-th component is equal to 1 if
vertex $i$ is in $S$, and 0 otherwise. The {\em stable set
polytope} of $G$, denoted by $STAB(G)$, is the convex hull of all
the incidence vectors of stable sets of $G$.

A system $Ax \leq b$ of linear inequalities is called a {\em defining linear system} for the stable set polytope of a graph $G$ if STAB($G$) =  $\{x \in \mathbb{R}^{|G|}: Ax \leq b\}$ holds and is $minimal$ if all inequalities are facet-defining; the (sub)graphs supporting such facet-defining inequalities are called facet-inducing graphs. As shown by Padberg, there are two types of inequalities that are necessary for all graphs: the nonnegativity constraints $-x_v \leq 0$ for all vertices $v \in V$ and the clique constraints $\sum_{v \in Q}x_v \leq 1$ for all inclusion-wise maximal cliques $Q \subseteq G$. Those two types of inequalities yield a minimal defining linear system for STAB($G$) if and only if $G$ is a perfect graph \cite{Chvatal1975}. In particular a well known result due to Chudnovsky et al. \cite{ChuRobSeyTho} states that a graph is perfect if and only if it contains no induced odd holes and no induced odd co-holes.

Finding such a system for the stable set polytope of general graphs is a difficult task, see e.g. \cite{GroLovSch1998,Pulleyblank1989,Schrijver1995,Schrijver2003}. That seems to be difficult also for graphs for which a maximum (weight) stable set can be computed in polynomial time: that is the case e.g. of claw-free graphs, though deep partial results have been stated, see e.g. \cite{EisOriStaVen2008,GalGenVen2011,GalSas1997,GilTro1981}.

However, it is known (see \cite{Chvatal1975}) how to construct the defining linear system for a graph obtained from substitution of a graph $G_2$ for a vertex $v$ of a graph $G_1$, provided that the defining linear system for STAB($G_1$) and for STAB($G_1$) are known. Furthermore, it is known (see e.g. \cite{DeSMos,Mosca2008}) that every non-trivial homogeneous set of a facet-inducing graph induces a facet-inducing graph as well. Then for any graph class $\cal{X}$ and for any $G \in \cal{X}$, an implicit description of STAB($G$) is given by the set $\cal{F_P(X)}$ of prime facet-inducing graphs of ${\cal X}$.

%The above remarks show that $\cal{F_P(X)}$ contains $K_2$ for any non-empty hereditary graph class $\cal {X}$ and that $\cal{F_P(X)}$ = $\{K_2\}$ if and only if $\cal{X}$ is a subclass of the class of perfect graphs.

In this paper we determine the set $\cal{F_P(X)}$ for the class $\cal {X}$ of ($P_6$,triangle)-free graphs and consequently give a complete description of STAB($G$) for every $G \in \cal {X}$ (these results are then extended to ($P_6$,paw)-free graphs). For that we combine, in analogy to the proof in \cite{DeSMos,Mosca2008} for some extensions of $P_4$-free graphs, different tools: Chv\'atal's results \cite{Chvatal1975}, properties of facet-inducing graphs due to Mahjoub \cite{Mahjoub1988}, a structure result on bipartite $P_6$-free graphs due to Fouquet, Giakoumakis, and Vanherpe \cite{FouGiaVan1999}, and intensively a structure result on non-bipartite prime ($P_6$,triangle)-free graphs due to Brandst\"adt, Klembt, and Mahfud \cite{BraKleMah2005}. Also we point out some peculiarities of the new facet-inducing graphs detected along this study with the help of a software.

The description of the stable set polytope for subclasses of triangle-free graphs may be of additional interest for the following reasons: (i) for triangle-free graphs the Maximum Stable Set Problem remains NP-hard \cite{Polja1974}; then the difficulty of describing their stable set polytope should be similar to that of the general case; (ii) for triangle-free graphs no substitution is possible (otherwise a triangle arises); then the difficulty of describing their stable set polytope is directly linked to that of detecting prime facet-inducing graphs.

%In addition we explicitly provide the defining linear system for three of the studied extensions of $P_4$-free graphs, showing that besides nonnegativity and clique constraints only rank constraints associated with extensions of odd holes and odd co-holes are required (see Section 3.2). $P_4$-free graphs are perfect, and our result shows that those extensions of $P_4$-free graphs belong to a superclass of perfect graphs, namely the rank-perfect graphs \cite{Wagler2002}.

\section{Basic notation and preliminary}

For any missing notation or reference let us refer to \cite{BraLeSpi1999,Pulleyblank1989}.

Let $G = (V,E)$ be a graph with $n$ vertices.

For any vertex-set $W \subseteq V$, let $N(W) = \{v \in V \setminus W : v$ is adjacent to some vertex of $W \}$; if
$W = \{w\}$, then let us write $N(w)$ instead of $N(\{w\})$; in particular $|N(v)|$ is the $degree$ of $v$. For any vertex-set $W \subseteq V$, let $G[W]$ denote the subgraph of $G$ induced by $W$. For convenience, in some part of the paper let us write $G - W$ instead of $G[V \setminus W]$.

A $clique$ of $G$ is a set of pairwise adjacent vertices of $G$.

For $q > 1$: let $K_q$ denote (the graph induced by) a clique with $q$ vertices, let $P_q$ denote
an induced path with $q$ vertices, and let $C_q$ denote an induced cycle with $q$ vertices.
Graph $K_3$ is also called $triangle$. A $paw$ has vertices $a,b,c,d$, and edges $ab,ac,bc,ad$, i.e.,
a paw is a one-vertex extension of a triangle.

Given a graph $F$, let us say that $G$ is $F$-$free$ if $G$ contains no induced subgraph isomorphic to $F$.
In particular, if $G$ is both $F_1$-free and $F_2$-free for some graphs $F_1,F_2$, then let us write $G$ is {\em $(F_1,F_2)$-free}.

The {\em complement} of $G$, denoted by {\em co-G}, is the graph
having the same vertices as $G$ and where two vertices are
adjacent in co-$G$ if and only if they are nonadjacent in $G$.

Two disjoint vertex-sets $X, Y \subseteq V$ have a $join$ (a {\em
co-join}) if each element of $X$ is adjacent (nonadjacent) to each
element of $Y$. A vertex $v \in V$ $distinguishes$ vertices $x,y
\in V$ if $(v,x) \in E$ and $(v,y) \not \in E$.
A vertex set $M \subseteq V$ is a $module$ of $G$ if no vertex
from $V \setminus M$ distinguishes two vertices from $M$. A module
is trivial if it is either the empty set, a one-vertex or the
entire vertex set $V$. Nontrivial modules are called {\em homogeneous sets}.

A graph is $prime$ if it contains only trivial modules.
The notion of modules is basic in the modular decomposition (or
substitution) of graphs (see e.g.
\cite{MohRad1984}). A homogeneous set $M$ is $maximal$ if no other
homogeneous set properly contains $M$. It is well known that in a
connected graph $G$ with connected complement co-$G$, the maximal
homogeneous sets are pairwise disjoint which means that every
vertex is contained in at most one homogeneous set. The existence
and uniqueness of the {\em modular decomposition tree} is based on
this property, and linear time algorithms were designed
to determine this tree (see e.g. \cite{McCSpi1999}). The tree
contains the vertices of the graph as its leaves, and the internal
nodes are of three types: they represent a join or a co-join
operation, or a prime graph.

\subsection{STAB(G) and modular composition}

Substitution of a graph $F$ for a vertex $v$ of a graph $G$ consists of
taking a disjoint union of $F$ and $G - \{v\}$, and
adding an edge between every vertex of $F$ and every vertex of
$G - \{v\}$ that was adjacent to $v$ in $G$; $G(v, F)$
denotes the graph obtained that way.

The following well known result of Chv\'atal (cf. Theorem 5.1 in
\cite{Chvatal1975}) gives the link between defining linear systems
of STAB($G$) and modular composition of graphs: if one knows a
defining linear system of STAB($G$) and a defining linear system
of STAB($F$), then one knows a defining linear system of
STAB($G(v,F)$).

\begin{theo} {\bf (\cite{Chvatal1975})}\label{theosubst}
Let $G_1=(V_1,E_1)$ and $G_2=(V_2,E_2)$ be graphs with $V_1\cap
V_2=\emptyset$. For $k=\{1,2\}$, let

$$\begin{array}{rl}
-x_u\leq 0                       &\quad (u\in V_k)\\
\displaystyle{\sum_{u\in V_k}a_{iu}x_u \leq b_i}  &\quad (i\in
J_k)
\end{array}$$

\smallskip
be a defining linear system of STAB($G_k$) (where $J_k$ is its inequalities index-set). Let $v$ be a node of
$G_1$ and let $G$ be the graph obtained from $G_1$ by substituting
$G_2$ for $v$. For each $i\in J_1$, set $a^+_{iv}=\max
\{a_{iv},0\}$. Then

$$\begin{array}{rl}
-x_u\leq 0                       &\quad (u\in V_2\cup (V_1-v))\\
a^+_{iv}\displaystyle{ \sum_{u\in V_2}a_{ju}x_u +b_j \sum_{u\in
V_1-v}a_{iu}x_u \leq b_ib_j}  &\quad (i\in J_1, j\in J_2)
\end{array}$$

\smallskip
is a defining linear system of STAB($G$). \qed
\end{theo}

An inequality is $valid$ for STAB($G$) if it is satisfied by each element of STAB($G$). A $face$ of STAB($G$) is the set $\{\bf{x}$ $\in$ STAB($G$) $: \sum_{i\in V}c_ix_i = b\}$ for some inequality $\sum_{i\in V}c_ix_i = b$ valid for $G$. A $facet$ of STAB($G$) is a maximal proper face of STAB($G$). An inequality $\sum_{i\in V}c_ix_i \leq
b$ is {\em facet-defining}, i.e., it belongs to a minimal defining linear system of STAB($G$), if and
only if it is valid for STAB($G$) and $\{\bf{x}$ $\in$ STAB($G$) $: \sum_{i\in V}c_ix_i = b \}$ is a facet of STAB($G$). Each facet of STAB($G$) is uniquely determined up to multipliers (cf. Theorem 3.16 of \cite{Pulleyblank1989}). In this sense, STAB($G$) admits a unique minimal defining linear system.

It is known that (see e.g. \cite{Pulleyblank1989}): if $\{\bf{x}$ $\in$ STAB($G$) $: \sum_{i\in V}c_ix_i = b \}$ is a facet of STAB($G$), then it is also a facet of STAB($G[W]$), where $W = \{i \in V: c_i \neq 0\}$. Then let us focus on the following kind of graphs.

A graph $G = (V,E)$ of $n$ vertices is {\em facet-inducing} if there exists a vector ${\bf c}=(c_1,\dots,c_n)^T$ with $c_i \neq 0$ for every $i$, and an integer $b$ such that the inequality $\sum_{i\in V}c_ix_i \leq
b$ is facet-defining for STAB($G$). Examples of facet-inducing graphs are the $C_k$ and the co-$C_k$ for $k = 2j+1$ and $j \geq 2$ \cite{Padberg1973,Padberg1974}. Let us call such an inequality $\sum_{i\in V}c_ix_i \leq
b$ as a {\em full facet} of STAB($G$). Actually one can assume that $b = 1$, according to Theorem 4 of \cite{DeSMos} which states that $b > 0$. Notice that STAB($G$) may have different full facets, see e.g. Figure 1 (c)-(d) of \cite{BarMah1994}. Then, for any facet-inducing graph $G$, let $\Phi (G)$ denote the set of full facets of STAB($G$).

Let us report the following result, one implication of which comes from Theorem 3.16 of \cite{Pulleyblank1989}.

\begin{theo}{\bf (\cite{DeSMos})}\label{theo facetinducing}
Let $G$ be a graph of $n$ vertices, $n>1$, and let
${\bf c}^T{\bf x} \leq b$ be an inequality valid for STAB($G$),
where ${\bf c}=(c_1,\dots,c_n)^T$ with $c_i \neq 0$ for every $i$.
Then the following statements are equivalent:

\smallskip
\noindent
(a) $G$ is facet-inducing and ${\bf c}^T{\bf x} \leq b$ is facet-defining for STAB($G$);

\smallskip
\noindent (b) there exists an $n\times n$ nonsingular matrix ${\bf
M}$, whose rows are incidence vectors of $n$ maximal stable sets
of $G$, such that ${\bf Mc} ={\bf b}$, where ${\bf b}$ denotes the
vector whose components are all equal to $b$. \qed
\end{theo}

{\bf Note:} As remarked above, one implication of Theorem \ref{theo facetinducing} comes from Theorem 3.16 of \cite{Pulleyblank1989}.
Actually, also the other implication of Theorem \ref{theo facetinducing} seems to be known earlier than \cite{DeSMos}, since it seems to be
applied (as a known fact) in some argument given in \cite{BarMah1994}.

%Theorem \ref{theo facetinducing} implies that: if a graph $G$ with $n$ vertices is facet-inducing, then $G$ contains $n$ maximal stable sets whose incidence vectors form a linearly independent set.

%The odd holes and the odd co-holes are well known examples of facet-inducing graphs. In particular: a minimal defining linear system for the stable set polytope of an odd hole $C_{2k+1}$ is given by nonnegativity constraints, clique constraints, and inequality $\sum _{i = 1,\ldots,2k+1} x_i \leq k$ by \cite{Padberg1973}; a minimal defining linear system for the stable set polytope of an odd co-hole co-$C_{2k+1}$ is given by nonnegativity constraints, clique constraints, and inequality $\sum _{i = 1,\ldots,2k+1} x_i \leq 2$ by \cite{Padberg1974}.

%Then, by Theorem \ref{theo facetinducing} and the definition of
%facet-inducing graphs, there exists a pair ({\bf M,c}), where {\bf
%M} is an $n \times n$ nonsingular matrix whose rows are the
%incidence vectors of $n$ maximal stable sets of $G$, and {\bf c}
%is an $n$-vector $(c_1,\ldots,c_n)$ with $c_i \neq 0$ for every
%$i$, such that: ($i$) inequality $\sum_{i\in V}c_ix_i \leq 1$ is facet-defining for STAB($G$); ($ii$) {\bf Mc} = {\bf 1}, where {\bf 1} denotes an $n$-vector whose entries are all equal to 1. Let us call ({\bf M, c}) a {\em pair associated} to $G$.

For every graph class ${\cal X}$, let ${\cal F(X)}$ denote the
class of all graphs in ${\cal X}$ that are facet-inducing. Clearly one has:

\begin{prop}\label{prop trivial}
For every graph class ${\cal X}$ and for every $G \in {\cal X}$, if one knows ${\cal F(X)}$, then through $\{\Phi(H) : H \in {\cal F(X)}\}$ one knows (explicitly) a linear defining system of STAB($G$).    \qed
\end{prop}

Let ${\cal F_P(X)}$ denote the class of all graphs in ${\cal F(X)}$
that are prime, and let ${\cal S(F_P(X))}$ denote the class of all
graphs that are obtained by possible repeated substitutions of any graph in
${\cal F_P(X)}$ for a vertex of any graph in ${\cal F_P(X)}$.
Clearly, ${\cal F_P(X)} \subseteq {\cal F(X)} \subseteq {\cal X}$
and ${\cal F_P(X)} \subseteq {\cal S(F_P(X))}$. Note that, if one
restricts attention to hereditary graph classes (i.e.,
defined by forbidding induced subgraphs), then
${\cal F_P(X)}$ contains $K_2$ for every non-empty hereditary graph class ${\cal X}$.
In particular, since $\cal{F(X)}$ is formed by cliques if and only if $\cal{X}$ is a subclass of the class of perfect graphs \cite{Chvatal1975},
$\cal{F_P(X)}$ = $\{K_2\}$ if and only if $\cal{X}$ is a subclass of the class of perfect graphs.

By Theorem \ref{theo facetinducing} one can prove the following proposition.

\begin{prop}{\bf (\cite{Mosca2008})}\label{green}
Let $G$ be a facet-inducing graph. Then every subgraph of $G$ induced by a
homogeneous set in $G$ is facet-inducing.       \qed
\end{prop}

By Theorem \ref{theosubst} and Proposition \ref{green} one obtains the following corollary, also quoted in \cite{DeSMos,Mosca2008}.

\begin{coro}\label{coro substitution}
For every graph class ${\cal X}$, ${\cal F(X)} = {\cal S(F_P(X))} \cap {\cal X}$.     \qed
\end{coro}

{\bf Proof.} By Theorem \ref{theosubst} one directly has ${\cal
S(F_P(X))} \cap {\cal X} \subseteq {\cal F(X)}$. Then let us prove that ${\cal F(X)} \subseteq {\cal
S(F_P(X))} \cap {\cal X}$. Let $G \in {\cal F(X)}$. By Proposition \ref{green} each node of the modular decomposition tree of $G$, different from those representing a join or a co-join operation, represents a prime facet-inducing graph. Then the assertion follows. \qed \\

Then one has:

\begin{prop}\label{prop trivial 2}
For every graph class ${\cal X}$ and for every graph $G \in {\cal X}$, if one knows ${\cal F_P(X)}$, then through $\{\Phi(H) : H \in {\cal F_P(X)}\}$ one knows (implicitly) a defining linear system of STAB($G$). \qed
\end{prop}

{\bf Proof.} If one knows ${\cal F_P(X)}$ and $\{\Phi(H) : H \in {\cal F_P(X)}\}$, then one can get ${\cal F(X)}$ by Corollary \ref{coro substitution} and $\{\Phi(H') : H' \in {\cal F(X)}\}$ by Theorem \ref{theosubst}, that is a linear defining system of STAB($G$) up to nonnegativity constraints.  \qed

\subsection{On the structure of facet-inducing graphs}

%In this subsection, let us report some structure results on facet-inducing graphs due to Mahjoub \cite{Mahjoub1988}, and mention a generalization of one of them due to a further result of Chv\'atal \cite{Chvatal1975}; furthermore, an observation from \cite{DeSMos} is reported, and two new observations are introduced.

Chv\'atal \cite{Chvatal1975} and Mahjoub \cite{Mahjoub1988} proved several results on the structure of facet-inducing graphs. In the sequel let us report just those results which will be used later. Also let us report an observation from \cite{DeSMos} and introduce two new observations.

First let us observe that every facet-inducing graph is connected: this fact, which is mentioned also in \cite{BarMah1994}, can be derived by Corollary \ref{coro substitution} and since every prime graph is connected.

The following lemma is an extract of the proof of Lemma 1 of \cite{Mahjoub1988} and is reported together with the proof, since such a proof idea/technique will be used later.

\begin{lemm}{\bf (\cite{Mahjoub1988})}\label{lemm Mahjoub}
Let $G$ be a facet-inducing graph and ${\bf c}^T{\bf t} \leq 1$ be a full facet of STAB($G$). Let $v$ be a vertex of $G$, of degree 2, with two non-adjacent neighbors $a,b$. Then $c_v \leq c_a$ and $c_v \leq c_b$.
\end{lemm}

{\bf Proof.} Let $S^*$ be the family of all maximal stable sets of $G$ whose incidence vectors ${\bf t}$ enjoy ${\bf c}^T{\bf t} = 1$. Then the only equations satisfied by all the incidence vectors of members of $S^*$ are positive multiplies of ${\bf c}^T{\bf t} = 1$. We claim that there exists a stable set $S_0 \in S^*$ such that $a \in S_0$ and $b \not \in S_0$. In fact, if this is not the case then for every stable set $S \in S^*$ the following holds: $a \in S$ implies $b \in S$; $a \not \in S$ implies $|S \cap \{v,b\}| = 1$. Thus $t_v + t_b = 1$ holds for all the incidence vectors of members of $S^*$, a contradiction. Let $S_0' = (S_0 \setminus \{a\}) \cup \{v\}$. Since $S'_0$ is also a stable set of $G$, we have $c_v \leq c_a$. Similarly, by symmetry, one obtains $c_v \leq c_b$.    \qed  \\

%\begin{lemm}{\bf (\cite{Mahjoub1988})}\label{lemma 1 Mahjoub}
%Let $G$ be a facet-inducing graph and ${\bf c}^T{\bf t} \leq 1$ be a full facet of STAB($G$). If $G$ contains a path ($pu,uv,vq$) such that $u$ and $v$ are of degree two, then $c_u = c_v$.    \qed
%\end{lemm}

%\begin{lemm}{\bf (\cite{Mahjoub1988})}\label{lemma 2 Mahjoub}
%Let $G$ be a facet-inducing graph, different to a hole and to a triangle. Let $p,q$ be nodes of $G$. Then at most one path in $G$ which joins $p$ and $q$ can have all internal nodes of degree two.     \qed
%\end{lemm}

%\begin{lemm}{\bf (\cite{Mahjoub1988})}\label{lemma 3 Mahjoub}
%Let $G$ be a facet-inducing graph. Then $G$ has no vertex of degree 1.     \qed
%\end{lemm}

%Then let us mention a generalization of the above result, which comes from Theorem 4.1 of \cite{Chvatal1975} (reported below) and the definition of facet-inducing graphs.

A $cutset$ of a graph $G = (V,E)$ is a subset $W$ of $V$ such that $G - W$ has more connected components that $G$. A {\em clique cutset} of $G$ is a cutset of $G$ which is also a clique of $G$.

Theorem 4.1 of \cite{Chvatal1975} shows that, given two graphs $G_1 = (V_1,E_1)$ and $G_2 = (V_2,E_2)$, if $(V_1 \cap V_2, E_1 \cap E_2)$ is a clique, then a defining linear system of $(V_1 \cup V_2,E_1 \cup E_2)$ is given by the union of linear defining systems of $G_1$ and $G_2$ respectively. Then by definition of facet-inducing graph one obtains:

%\begin{theo}{\bf (\cite{Chvatal1975})}\label{theo clique cutset}
%Given two graphs $G_1 = (V_1,E_1)$ and $G_2 = (V_2,E_2)$, if $(V_1 \cap V_2, E_1 \cap E_2)$ is a clique, then a defining linear system of $(V_1 \cup V_2,E_1 \cup E_2)$ is given by the union of linear defining systems of $G_1$ and $G_2$ respectively.   \qed
%\end{theo}

\begin{theo}{\bf (\cite{Chvatal1975})}\label{theo clique cutset}
Every facet inducing-graph has no clique cutset.    \qed
\end{theo}

Let us say that a vertex $u$ of a facet-inducing graph $G$ with $n$ vertices, with $n>1$,
is {\em critical} for $G$ if there exists a matrix ${\bf M}$ according to Theorem \ref{theo facetinducing}
such that the column of ${\bf M}$ corresponding to vertex $u$ has a unique entry equal to 1.

\begin{obse}\label{obse critical1} {\bf (\cite{DeSMos})}
Let $G$ be a facet-inducing graph. If $v$ is a critical vertex for
$G$, then $G - \{v\}$ is facet-inducing.      \qed
\end{obse}

Let us conclude this subsection by introducing two observations.

Let us say that a subgraph $G[H]$ of a graph $G$ is {\em repeating} for $G$ if for each maximal stable set $S$ of $G$ such that $S \cap H \neq \emptyset$ one has that $S \cap H$ is maximal for $G[H]$.

\begin{obse}\label{obse repeating}
Let $G$ be a facet-inducing graph and $G[H]$ be a repeating subgraph of $G$. Then $G[H]$ contains $|H|$ maximal stable sets whose incidence vectors form a linearly independent set.
\end{obse}

{\bf Proof.} Since $G$ is facet-inducing, there exists a matrix ${\bf M}$ according to Theorem \ref{theo facetinducing}. Since ${\bf M}$ is nonsingular, the submatrix ${\bf M'}$ of ${\bf M}$ formed by the columns corresponding to the vertices of $G[H]$ has rank $|H|$. Then since $G[H]$ is repeating for $G$, the rows of ${\bf M'}$ are incidence vectors of maximal stable sets of $G[H]$, and the assertion follows.    \qed

\begin{obse}\label{obse degree 2}
Let $G$ be a facet-inducing graph and ${\bf c}^T{\bf t} \leq 1$ be a full facet of STAB($G$). Let $v$ be a vertex of $G$, of degree 2, with two non-adjacent neighbors $a,b$. Then there exists a maximal stable set $S$ of $G$ with incidence vector ${\bf t'}$ such that $a,b \in S$, and ${\bf c}^T{\bf t'} = 1$.
\end{obse}

{\bf Proof.} Let $S^*$ be the family of all maximal stable sets of $G$ whose incidence vectors ${\bf t}$ are such that ${\bf c}^T{\bf t} = 1$. Then the only equations satisfied by all the incidence vectors of members of $S^*$ are positive multiplies of ${\bf c}^T{\bf t} = 1$. Assume to the contrary that there exists no maximal stable set in $S^*$ containing both $a$ and $b$. Then each maximal stable set in $S^*$ contains exactly one vertex from $\{a,v,b\}$. Then $t_{a} + t_{v} + t_{b} = 1$ holds for all maximal stable sets of $S^*$, a contradiction.   \qed

\section{Some observations on facet-inducing graphs which are either triangle-free or $P_6$-free}

Let us consider triangle-free graphs.

\begin{obse}\label{obse triangle-free}
Let ${\cal X}$ be the class of triangle-free graphs. Then ${\cal F(X)} = {\cal F_P(X)} = {\cal S(F_P(X))}$.
\end{obse}

{\bf Proof.} In fact every prime facet-inducing graph contains at least one edge; then to avoid a triangle one has ${\cal F_P(X)} = {\cal S(F_P(X))}$; then by Corollary \ref{coro substitution} the assertion follows.   \qed  \\

Let us consider $P_6$-free graphs.

Let us report a result due to Fouquet, Giakoumakis, and Vanherpe \cite{FouGiaVan1999} (see also \cite{GiaVan2003}). Given a bipartite graph $G=(S_1 \cup S_2,F)$, the $bi$-$complemented$ graph $\overline{G}^{bip}$ is the graph having
the same vertex set $S_1 \cup S_2$ as $G$ while its edge set is equal to $(S_1 \times S_2) \setminus F$.

\begin{theo}[\cite{FouGiaVan1999}]\label{P6 bipartite structure}
Let $G=(S_1 \cup S_2,F)$ be a connected bipartite $P_6$-free
graph. Then one of the following cases occurs:
\begin{enumerate}
\item[$(i)$]   $\overline{G}^{bip}$ is disconnected;
\item[$(ii)$]  there exist $S^*_1 \subseteq S_1$ and $S^*_2 \subseteq
S_2$ such that $G[S^*_1 \cup S^*_2]$ is complete bipartite, and
$(S_1 \setminus S^*_1) \cup (S_2 \setminus S^*_2)$ is an
independent set.       \qed
\end{enumerate}
\end{theo}

\begin{lemm}\label{lemm P6 maximal stable sets}
Let $G=(S_1 \cup S_2,F)$ be a connected bipartite $P_6$-free
graph, different to $K_2$. Then $G$ contains less than $|S_1 \cup S_2|$ maximal stable sets.
\end{lemm}

{\bf Proof.} For every subset $U$ of $V = S_1 \cup S_2$, let $m(U)$ be the number of maximal stable sets contained in $G[U]$.
Referring to Theorem \ref{P6 bipartite structure}, let us consider the following cases.

Assume that case ($i$) occurs. Then let $K_1,\ldots,K_t$ be the vertex sets of the connected components of $\overline{G}^{bip}$. Then $m(V) = m(K_1) + \ldots + m(K_t) - 2t + 2$, where $2t$ is the number of the sides of each $K_i$ ($i = 1,\ldots,t$) and 2 is the number of the sides of $V$ (which are $S_1$ and $S_2$). Thus since $t > 1$, if $m(K_i) \leq |K_i|$ for $i = 1,\ldots,t$, then the lemma follows.

Assume that case ($ii$) occurs. If $S_1 \setminus S^*_1 = S_2 \setminus S^*_2 = \emptyset$, then $m(V) = 2$, i.e., the lemma follows since $G$ is not a $K_2$. If $S_1 \setminus S^*_1 = \emptyset$ and $S_2 \setminus S^*_2 \neq \emptyset$, then $m(V) = m(S_1^* \cup (S_2 \setminus S^*_2))$ (in fact there is just one maximal stable set of $G$ containing vertices of $S^*_2$, that is $S_2$). If $S_1 \setminus S^*_1 \neq \emptyset$ and $S_2 \setminus S^*_2 = \emptyset$, then similarly $m(V) = m(S_2^* \cup (S_1 \setminus S^*_1))$. If $S_1 \setminus S^*_1 \neq \emptyset$ and $S_2 \setminus S^*_2 \neq \emptyset$, then similarly $m(V) = m(S_1^* \cup (S_2 \setminus S^*_2)) + m(S_2^* \cup (S_1 \setminus S^*_1)) - 1$ (since the maximal stable set $(S_1 \setminus S^*_1) \cup (S_2 \setminus S^*_2)$ has been considered twice). Thus, concerning the last three cases, if $m(S_1^* \cup (S_2 \setminus S^*_2)) \leq |S_1^* \cup (S_2 \setminus S^*_2)|$ and $m(S_2^* \cup (S_1 \setminus S^*_1)) \leq |S_2^* \cup (S_1 \setminus S^*_1)|$, then the lemma follows.

By Theorem \ref{P6 bipartite structure} one can repeatedly iterate the above arguments for each $K_i$ ($i = 1,\ldots,t$), for $S_1^* \cup (S_2 \setminus S^*_2)$ and for $S_2^* \cup (S_1 \setminus S^*_1)$, until to reach subgraphs of $G$, say $H$, such that $H$ either is empty or is an edge, that is such that $m(H) \leq |H|$. Then the lemma follows.   \qed

\begin{defi} A subgraph $G[H]$ with no isolated vertices of a graph $G$ is a {\em bi-module} of $G$ if:

\begin{enumerate}
\item[$(i)$]   $G[H]$ is bipartite, i.e., $G[H] = (H_1 \cup H_2,F)$;
\item[$(ii)$]  $H_1$ ($H_2$) is a module of $G - H_2$ (of $G - H_1$).
\end{enumerate}

\end{defi}

According to the above definition, each edge of a graph $G$ is trivially a bi-module of $G$.

\begin{lemm}\label{lemm bi-module}
Let $G$ be a facet-inducing $P_6$-free graph. Then each bi-module of $G$ is an edge of $G$.
\end{lemm}

{\bf Proof.} Let $G[H]$ be a bi-module of $G$. By definition of bi-module and since $G[H]$ has no isolated vertices, $G[H]$ is a repeating subgraph of $G$. Then by Observation \ref{obse repeating} $G[H]$ contains $|H|$ maximal stable sets whose incidence vectors form a linearly independent set.

If $G[H]$ is connected, then by Lemma \ref{lemm P6 maximal stable sets} and since $G[H]$ is bipartite $P_6$-free, $G[H]$ contains $|H|$ maximal stable sets whose incidence vectors form a linearly independent set if and only if $G[H]$ is an edge of $G$. If $G[H]$ is not connected, then, since each connected component of a bi-module is a bi-module as well, by the previous sentence $G[H]$ is formed by (at least two) disjoint edges: then $G[H]$ does not contain $|H|$ maximal stable sets whose incidence vectors form a linearly independent set, that is, this case in not possible.   \qed \\

%Proof of (ii). By statement (i) each bi-module of $G$ is an edge of $G$. Thus if $G$ contains a blue bi-module, then $G$ has a vertex of degree 1, which contradicts Lemma \ref{lemma 3 Mahjoub}.  \qed

Let us conclude this section by pointing out a class of ($P_6$,triangle)-free graphs which are not facet-inducing $-$ this will be useful later.

A graph $G$ is

\begin{itemize}
\item {\em matched co-bipartite} if $G$ is partitionable into two cliques $C_1, C_2$ with $|C_1| = |C_2|$ or $|C_1| = |C_2| - 1$ such that the edges between $C_1$ and $C_2$ are a matching and at most one vertex in $C_1$ and $C_2$ is not covered by the matching;
\item {\em co-matched bipartite} if it is the complement of a matched co-bipartite graph.
\end{itemize}

Notice that every co-matched bipartite graph is not facet-inducing, since it contains not enough maximal stable sets, according to Theorem \ref{theo facetinducing}.

Let us call $ferry$ a graph $F = (X \cup Y \cup Z, E)$ where $X,Y,Z$ are respectively stable sets, $X = \{x_0,x_1,\ldots,x_m\}$, $Y = \{y_0,y_1,\ldots,y_m\}$, $x_0$ may not exist and dominates $Y$, $y_0$ may not exist and dominates $X$, $x_i$ is adjacent to each vertex of $Y$ except for $y_i$ for $i = 1,\ldots,m$ (i.e., $X \cup Y$ induce a co-matching bipartite graph), $Z = \{z_1,\ldots,z_l\}$ with $l \leq m$ and, for every $i = 1,\ldots,l$, $z_i$ is of degree 2 and is adjacent to $x_i$ and $y_i$ $-$ see Figure 1.

\begin{figure}
\centering
\includegraphics[width=\textwidth]{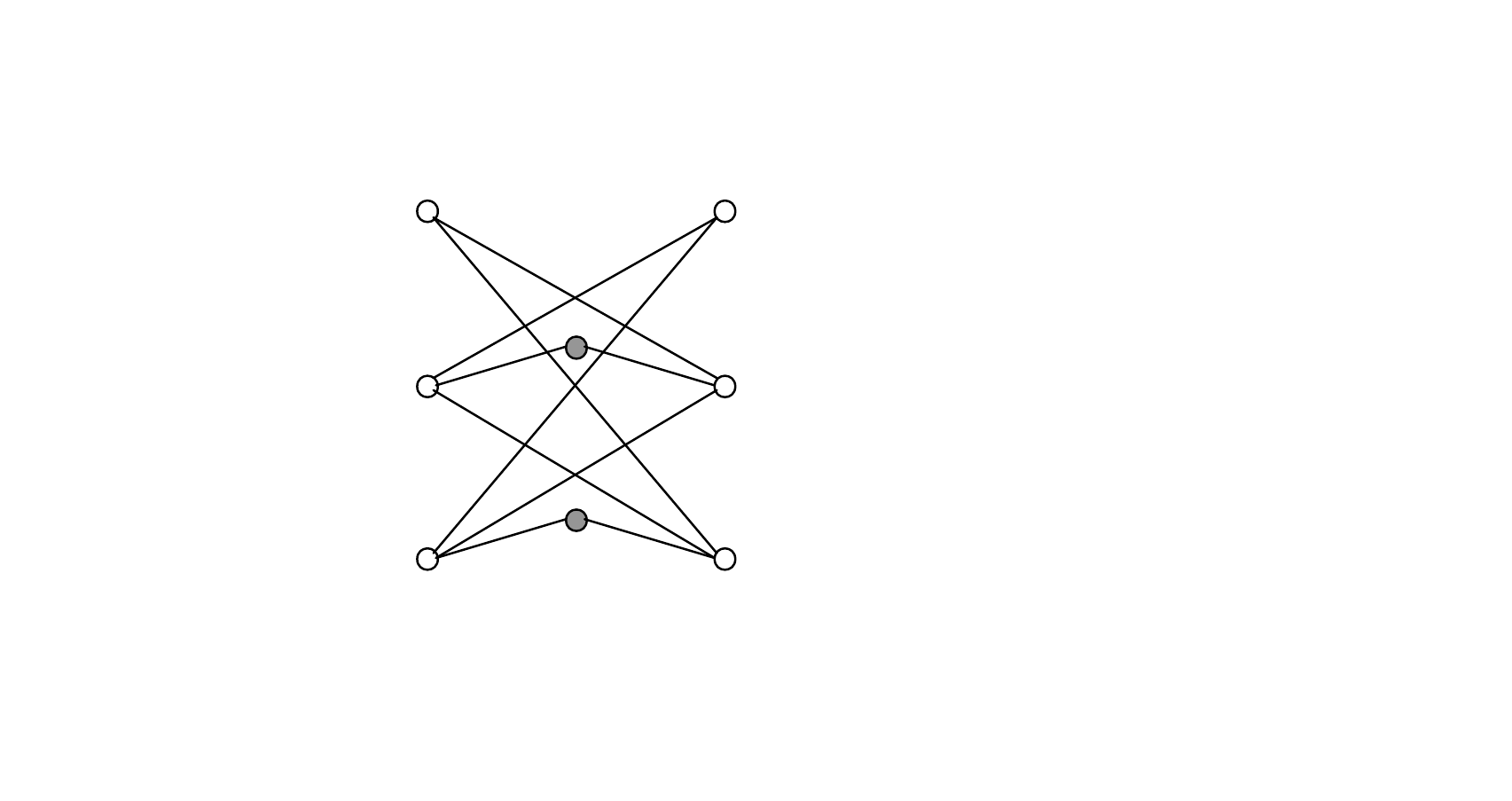}
\caption{A ferry with $m$ = 3 and $l$ = 2 (without $x_0$ and $y_0$)}
\end{figure}

\begin{lemm}\label{F_m}
Every ferry is not facet-inducing.
\end{lemm}

{\bf Proof.} Assume to the contrary that a ferry $F = (X \cup Y \cup Z, E)$ is facet-inducing.
Let us prove the lemma only for the case in which $x_0$ and $y_0$ do not exist; the case in which they exist can be similarly treated. Then $Z \neq \emptyset$, since co-matched bipartite graphs are not facet-inducing. Since $F$ is facet-inducing, let ${\bf c}^T{\bf t} \leq 1$ be a full facet of STAB($F$). Let $S^*$ be the family of all maximal stable sets of $F$ such that their incidence vectors ${\bf t}$ enjoy ${\bf c}^T{\bf t} = 1$. Then the only equations satisfied by all the incidence vectors of members of $S^*$ are positive multiplies of ${\bf c}^T{\bf t} = 1$. For brevity let us say that the members of $S^*$ are $green$ sets.

Let us observe that a maximal stable set of $F$ may be just of three types: $side$ if it is either $X$, or $Y$, or $Z$; $cross$ if it is formed by $x_i,y_i$ and all $z_j$'s with $j \neq i$; $balance$ if it is formed by a subset of $m$ vertices piked up either in both $X$ and $Z$ (balance ($X,Y$)) or in both $Y$ and $Z$ (balance ($Y,Z$)).

For any vertex $v$ of $F$, let us write $c(v)$ instead of $c_v$.

{\bf Claim 1} {\em $c(z_i) \leq c(x_i)$ and $c(z_i) \leq c(y_i)$ for every $i = 1,\ldots,m$}

{\bf proof.} It follows by Lemma \ref{lemm Mahjoub}.   \qed

{\bf Claim 2} {\em $X$ and $Y$ are green}

{\bf proof.} Assume to the contrary that $X$ is not green. Let $x_i \in X$ be such that $z_i$ does exist. Notice that $x_i$ is contained in at least one green balance ($X,Z$): in fact, otherwise, $x_i$ contained in at most one green maximal stable set (i.e., in a green cross); this implies that $x_i$ is critical for $F$; then by Observation \ref{obse critical1}, $F - \{x_i\}$ is facet-inducing; but $F - \{x_i\}$ contains a vertex of degree 1, i.e., vertex $z_i$, a contradiction to Theorem \ref{theo clique cutset}. %Lemma \ref{lemma 3 Mahjoub}.
Then let $S$ be a green balance ($X,Z$) containing $x_i$. Let $S_x$ (let $S_z$) denote the set of indices $i$ such that $x_i \in S$ ($z_i \in S$).
Then $\sum_{i \in S_x} c(x_i) +  \sum_{j \in S_z} c(z_j) = 1$. By Claim 1 and since inequality ${\bf c}^T{\bf t} \leq 1$ is valid for STAB($F$), one has $c(x_j) = c(z_j)$ for every $j \in S_z$. Then $\sum_{i = 1,\ldots,m} c(x_i) = 1$, i.e., $X$ is green. Similarly, by symmetry, one obtains that $Y$ is green as well.   \qed

{\bf Claim 3} {\em $Z$ is not green}

{\bf proof.} Assume to the contrary that $Z$ is green. Then $\sum_{1,\ldots,m} c(z_i) = 1$. On the other hand, one has $c(x_1) + c(y_1) + \sum_{2,\ldots,m} c(z_i) \leq 1$, since inequality ${\bf c}^T{\bf t} \leq 1$ is valid for STAB($F$). This is a contradiction since by Claim 1 $c(z_1) \leq c(x_1)$.   \qed

{\bf Claim 4} {\em Each cross is green}

{\bf proof.} Assume to the contrary that the cross formed by $x_i,y_i$ and all $z_j$'s with $j \neq i$ is not green. Then there exists no green maximal stable set containing both $x_i$ and $y_i$. This contradicts Observation \ref{obse degree 2}.   \qed
%Then each green maximal stable set contains exactly one vertex from $\{x_i,z_i,y_i\}$. Then inequality $t_{x_i} + t_{z_i} + t_{y_i} = 1$ is valid for all green maximal stable sets, a contradiction.      \qed

Let $\tilde{X} = \{x_i \in X: c(x_i) > c(z_i)\}$ and $\tilde{Y} = \{y_i \in Y: c(y_i) > c(z_i)\}$.

{\bf Claim 5} {\em $|\tilde{X}| = 1$ and $|\tilde{Y}| = 1$}

{\bf proof.} Let us consider only $\tilde{Y}$. The case of $\tilde{X}$ can be similarly treated by symmetry. Let us observe that $|\tilde{Y}| \geq 1$: in fact $|\tilde{Y}| = 0$ implies (by Claim 1) that $Z$ is green, a contradiction to Claim 3.

Then assume by contradiction that $|\tilde{Y}| > 1$. Without loss of generality let $\tilde{Y} = \{y_1,\ldots,y_q\}$ with $1 < q \leq m$. Let $G[H]$ be the graph induced by $\{x_1,\ldots,x_q\} \cup \{z_1,\ldots,z_q\}$. By definition of $\tilde{Y}$, for each green maximal stable set $S$ one has either $S \supseteq \tilde{Y}$ or $S \cap \tilde{Y} = \emptyset$. Then $G[H]$ is a repeating subgraph of $G$: then by Observation \ref{obse repeating} $G[H]$ contains $|H|$ maximal stable sets whose incidence vectors form a linearly independent set. This is not possible since $G[H]$ is formed by (at least two) disjoint edges.       \qed

Let us conclude the proof of the lemma. By Claim 5, without loss of generality let $\tilde{X} = \{x_1\}$. By Claim 2, $\sum_{i=1,\ldots,m}c(x_i) = 1$; then by definition of $\tilde{X}$, $c(x_1) + \sum_{i=2,\ldots,m}c(z_i) = 1$. On the other hand by Claim 4, $c(x_1)+c(y_1)+ \sum_{i=2,\ldots,m}c(z_i) = 1$, a contradiction. \qed

\section{Structure of prime ($P_6$,triangle)-free graphs from \cite{BraKleMah2005}}\label{section Germany}

In this section for the sake of completeness let us report those results from \cite{BraKleMah2005} which describe the structure of prime ($P_6$,triangle)-free graphs.

Throughout this section let $G = (V,E)$ be a non-bipartite prime ($P_6$,triangle)-free graph. For a subgraph $H$ of $G$, a vertex not in $H$ is a {\em k-vertex of H} (or {\em for H}) if it has exactly $k$ neighbors in $H$. We say that {\em H has no k-vertices} if there is no $k$-vertex for $H$.

Since $G$ is not bipartite, $G$ must contain an odd cycle of length at least 5. In particular, since $G$ is $P_6$-free, $G$ must contain a $C_5$, say, $C$ with vertices $v_1,\ldots,v_5$ and edges $\{v_i,v_{i+1}\}, i \in \{1,\ldots,5\}$ (throughout this section, all index arithmetic with respect to a $C_5$ is done modulo 5). Obviously, in a triangle-free graph, a $C_5$ $C$ has no 3-, 4- and 5-vertex, and 2-vertices of $G$ are have nonconsecutive neighbors in $C$. Let $X$ denote the set of $0$-vertices of $C$, and for $i = 1,\ldots,5$, let $Y_i$ denote the set of 1-vertices of $C$ being adjacent to $v_i$, and let $Z_{i,i+2}$ denote the set of 2-vertices of $C$ being adjacent to $v_i$ and $v_{i+2}$.

Moreover, let $Y = Y_1 \cup \ldots \cup Y_5$ and $Z = Z_{1,3} \cup Z_{2,4} \cup Z_{3,5} \cup Z_{4,1} \cup Z_{5,2}$. Obviously, $\{v_1,\ldots,v_5\} \cup X \cup Y \cup Z$ is a partition of $V$.

The following result comes from Section 3 of \cite{BraKleMah2005}.

\begin{lemm}{\bf (\cite{BraKleMah2005})}\label{lemm 1 Germany}
The following facts holds (for $i = 1,\ldots,5$):

\begin{enumerate}
\item[$(i)$] $X$ is a stable set;
\item[$(ii)$] $X$ has a co-join to $Y$;
\item[$(iii)$] $Y_i$ and $Z_{i,i+2}$ are stable sets;
\item[$(iv)$] $Y_i$ has a join to $Y_{i+2} \cup Y_{i+3}$, and a co-join to $Y_{i+1} \cup Y_{i+4}$;
\item[$(v)$] $Z_{i,i+2}$ has a co-join to $Z_{i,i+3} \cup Z_{i+2,i+4}$;
\item[$(vi)$] $Y_i$ has a co-join to $Z_{i,i+2} \cup Z_{i,i+3}$;
\item[$(vii)$] vertices in $Y_i$ can only be distinguished by vertices in $Z_{i-1,i+1}$.  \qed
\end{enumerate}

\end{lemm}

For the following we need the following notations:

$Z^0_{i,i+2}:= \{x: x \in Z_{i,i+2}$ and $x$ has a nonneighbor in $Z_{i-1,i+1}$ or in $Z_{i-1,i+1} \}$ for $i \in \{1,\ldots,5\}$, and let

$Z_0 = \cup_{i=1}^5 Z^0_{i,i+2}$.

Let $X_0$ denote the set of 0-vertices being adjacent to a vertex in $Z_0$ and let

$G_0:= G[X_0 \cup Z_0]$.

The following result comes from Section 4 of \cite{BraKleMah2005}.

\begin{lemm}{\bf (\cite{BraKleMah2005})}\label{lemm G_0 Germany}
One of the following cases occurs: $G_0$ is
\begin{enumerate}
\item[$(i)$] with no vertices;
\item[$(ii)$] formed by at most five vertices, with $|Z^0_{i,i+2}| \leq 1$ for each $i = 1,\ldots,5$; in this case, $|X_0| \leq 1$ and $X_0$ has a join to $Z_0$.
\item[$(ii)$] a co-matched bipartite, namely ($Z^0_{i+1,i+3} \cup Z^0_{i+2,i+4}$,$F'$); in this case each vertex of $X_0$ is adjacent to exactly a pair of nonadjacent vertices $a,b$ with $a \in Z^0_{i+1,i+3}$ and $b \in Z^0_{i+2,i+4}$; in other words, $Z^0_{i+1,i+3} \cup Z^0_{i+2,i+4} \cup X_0$ induces a ferry;
\item[$(iv)$] the disjoint union of two co-matched bipartite graphs, namely ($Z^0_{i+1,i+3} \cup Z^0_{i+2,i+4}$,$F'$) and ($Z^0_{i,i+3} \cup Z^0_{i+1,i+4}$,$F''$); in this case, $X_0 = \emptyset$. \qed
\end{enumerate}
\end{lemm}

Let $Z^1_{i,i+2}:= Z_{i,i+2} \setminus Z^0_{i,i+2}$ and $Z_1 := Z \setminus Z_0$. For $i \in \{1,\ldots,5\}$, let $X_i$ denote the set of 0-vertices being adjacent to $Z^1_{i-1,i+1}$. Now, if for $i \in \{0,1,\ldots,5\}$, $X_i$ is trivial, we will omit the single vertex in $X_i$, i.e.: if $X_i$ is nontrivial, then $X'_i = X_i$; if $X_i$ is trivial, then $X'_i = \emptyset$.

For $i \in \{1,\ldots,5\}$, let $B_i := G[X'_i \cup Y_i \cup Z^1_{i-1,i+1}]$. By Lemma \ref{lemm 1 Germany}, $X \cup Y_i$ is a stable set, and thus, $B_i$ is bipartite. Let $X_T$ denote the union of trivial $X_i$, $i \in \{0,1,\ldots,5\}$.

The {\em basic subgraphs} in $G$ are the subgraphs $G_0$ and $B_i$, $i \in \{1,\ldots,5\}$.

\begin{lemm}{\bf (\cite{BraKleMah2005})}\label{lemm 2 Germany}

The vertex sets $X'_0,Z_0$ of $G_0$ and the vertex sets $X'_i,Y'_i,Z^1_{i-1,i+1}$ of $B_i$, $i \in \{1,\ldots,5\}$, define a partition of $V \setminus (\{v_1,\ldots,v_5\} \cup X_T)$.       \qed

\end{lemm}

Recall that by Lemma \ref{lemm 1 Germany} (vii), vertices in $Y_i$ can only be distinguished by vertices in $Z_{i-1,i+1}$. Thus, every vertex in $Z_{i-1,i+1}$ has either a join or a co-join to $Y_{i+2}$ ($Y_{i+3}$, respectively).

Let $Z_{i-1,i+1;00}$ ($Z_{i-1,i+1;01}$, $Z_{i-1,i+1;10}$, $Z_{i-1,i+1;11}$) be the set of 2-vertices in $Z_{i-1,i+1}$ having a co-join to $Y_{i+2}$ and $Y_{i+3}$ (having a co-join to $Y_{i+2}$ and a join to $Y_{i+3}$, having a join to $Y_{i+2}$ and a co-join to $Y_{i+3}$, having a join to $Y_{i+2}$ and a join to $Y_{i+3}$). Moreover, let $Z^a_{i-1,i+1;bc}$ = $Z^a_{i-1,i+1} \cap Z_{i-1,i+1;bc}$, $a \in \{0,1\}$, $bc \in \{00,01,10,11\}$.

The {\em basic vertex subsets} of $G$ are $X'_0,X'_1,\ldots,X'_5$, $Y_1,\ldots,Y_5$, and $Z^a_{i-1,i+1;bc}$, $a \in \{0,1\}$, $bc \in \{00,01,10,11\}$.

\begin{theo}{\bf (Structure Theorem \cite{BraKleMah2005})}\label{theo Germany}

For all pairs of basic vertex subsets $U,W$ from different basic subgraphs, $U$ has either a join or a co-join to $W$. \qed
\end{theo}

\section{Structure of prime facet-inducing ($P_6$,triangle)-free graphs}

In this section let us describe the structure of prime facet-inducing ($P_6$,triangle)-free graphs.

Throughout this section let $G = (V,E)$ be a non-bipartite prime facet-inducing ($P_6$,triangle)-free graph. This is motivated from the fact that, since bipartite graphs are perfect, every (non-empty) bipartite prime facet-inducing graph is a $K_2$. Then let us adopt the notation of Section \ref{section Germany}.

%First let us introduce some results on subgraphs $B_i$ for $i = 1,\ldots,5$.

%\begin{prop}\label{prop 1}
%If $Y_i \neq \emptyset$, $Y_{i+2} \cup Y_{i+3} \neq \emptyset$, then $T$ is partitioned into $T_1=\{t \in T: t$ dominates $Y_i\}$ and $T_2=\{t \in T: t$ dominates $Y_{i+2} \cup Y_{i+3}\}$.
%\end{prop}
%{\bf Proof.} Since $G$ is $P_6$-free, $Y_i$ has a join to $Y_{i+2} \cup Y_{i+3}$. To avoid a $P_6$, each vertex of $T$ either dominates $Y_i$ or $Y_{i+2} \cup Y_{i+3}$, not both since $G$ is triangle-free    \qed

Let us consider $X'_i$ and $Y_i$.

\begin{lemm}\label{lemm X'_i}

$X'_i = \emptyset$, for $i = 1,\ldots,5$.

\end{lemm}

{\bf Proof.} Assume to the contrary that $X'_i \neq \emptyset$ for some $i = 1,\ldots,5$. For brevity let us write $T = Z^1_{i-1,i+1}$. Let $T' = T \cap N(X'_i)$: then $T' \neq \emptyset$, by definition of $X'_i$ and since $X'_i \neq \emptyset$.

{\bf Claim 1} {\em $X'_i$ is a module of $G - T'$}.

{\bf proof.} Assume by contradiction that there is a vertex $d$ of $G - T'$ distinguishing two vertices $x_1,x_2$ of $X'_i$. By Lemma \ref{lemm 1 Germany} and since $X'_i$ has a co-join to $Z_{i,i+2} \cup Z_{i,i+3}$ (otherwise a triangle arises with a vertex of $T'$), one has that $d \in Z_{i+1,i+3} \cup Z_{i+2,i+4}$, say $d \in Z_{i+1,i+3}$ being adjacent to $x_1$ and nonadjacent to $x_2$, without loss of generality. If $x_1$ and $x_2$ share a neighbor $t$ in $T$, then $x_2,t,x_1,d,v_{i+3},v_{i+2}$ induce a $P_6$. Otherwise, say $x_1$ is adjacent to $t_1 \in T$ and $x_2$ is adjacent to $t_2 \in T$, one has that $x_2,t_2,v_{i+4},t_1,x_1,d$ induce a $P_6$, contradiction.   \qed

{\bf Claim 2} {\em $T'$ has a join to $Y_{i+2} \cup Y_{i+3}$}.

{\bf proof.} In fact let $t \in T'$ be adjacent to $x \in X'_i$. Then $t$ is adjacent to each vertex $y \in Y_{i+2}$, otherwise $x,t,v_{i+4},v_{i+3},v_{i+2},y$ induce a $P_6$. Similarly, by symmetry, $t$ is adjacent to each vertex of $Y_{i+3}$.   \qed

To conclude the proof of the lemma let us consider the following cases.

{\bf Case 1} $T$ has a co-join to $Y_i$

Then by Claims 1 and 2, $T'$ and $X'_i$ form a bi-module of $G$. By Lemma \ref{lemm bi-module} they are an edge of $G$, i.e. $|X'_i| = 1$, a contradiction to the definition of $X'_i$.

{\bf Case 2} $T$ has a not a co-join to $Y_i$

This case means that $Y_i \neq \emptyset$. Then $Y_{i+2} \cup Y_{i+3} = \emptyset$: in fact otherwise by Lemma \ref{lemm 1 Germany} and Claim 2 a triangle arises with a vertex of $Y_i$ and a vertex of $T'$.

Let us prove that $T \cup \{v_i\}$ and $X'_i \cup Y_i$ form a bi-module of $G$. From one hand, $T \cup \{v_i\}$ is clearly a module of $G - (X'_i \cup Y_i)$. On the other hand, assume by contradiction that there is $d \in V \setminus (T \cup \{v_i\})$ distinguishing two vertices $y,x$ of $X'_i \cup Y_i$. By Theorem \ref{theo Germany}, Claim 1 and since $Y_{i+2} \cup Y_{i+3} = \emptyset$, one has that $x \in X'_i$, $y \in Y_i$, $d \in Z_{i+1,i+3} \cup Z_{i+2,i+4}$, say $d \in Z_{i+1,i+3}$ without loss of generality. Let $t \in T$ be a neighbor of $x$. If $d$ is adjacent to $x$ and nonadjacent to $y$, then: if $y$ is nonadjacent to $t$, then $y,v_i,v_{i+4},t,x,d$ induce a $P_6$; if $y$ is adjacent to $t$, then $v_i,y,t,x,d,v_{i+3}$ induce a $P_6$. If $d$ is adjacent to $y$ and nonadjacent to $x$, then: if $y$ is nonadjacent to $t$, then $y,d,v_{i+3},v_{i+4},t,x$ induce a $P_6$; if $y$ is adjacent to $t$, then $v_{i+2},v_{i+3},d,y,t,x$ induce a $P_6$, a contradiction.

Then $T \cup \{v_i\}$ and $X'_i \cup Y_i$ form a bi-module of $G$. By Lemma \ref{lemm bi-module} they are an edge of $G$, i.e., $T = \emptyset$ and consequently $X'_i = \emptyset$, a contradiction.     \qed

\begin{lemm}\label{lemm B_i}

$B_i$ is a stable set, for $i = 1,\ldots,5$.

\end{lemm}

{\bf Proof.} Assume to the contrary that $B_i$ is not a stable set for some $i = 1,\ldots,5$. For brevity let us write $T = Z^1_{i-1,i+1}$. By Lemma \ref{lemm X'_i}, $X'_i = \emptyset$: then $B_i$ is formed by $T$ and $Y_i$.

First assume that the possible trivial $X_i$ does not exist. Then by Theorem \ref{theo Germany}, $T \cup \{v_i\}$ and $Y_i$ form a bi-module of $G$. Then by Lemma \ref{lemm bi-module}, they are an edge of $G$ (i.e., $T = \emptyset$), which implies that $B_i$ is a stable set, a contradiction.

Then assume that the possible trivial $X_i$ does exist, say vertex $x$. Since $G$ is triangle-free, $x$ is adjacent to no vertex of $Z_{i,i+2} \cup Z_{i,i+3}$. Then one can apply an argument similar to that of Lemma \ref{lemm X'_i}, with $\{x\}$ instead of $X'_i$, to get a contradiction (notice that Case 1 is not possible since $B_i$ is not a stable set).     \qed

\begin{lemm}\label{lemm Y_i}

The following facts hold for $i = 1,\ldots,5$:

\begin{enumerate}
\item[$(i)$] $Z_{i-1,i+1}$ has a co-join to $Y_i$;
\item[$(ii)$] $|Y_i| \leq 1$.
\end{enumerate}
\end{lemm}

{\bf Proof.} Proof of (i). It follows by Theorem \ref{theo Germany} and by Lemma \ref{lemm B_i}. Proof of (ii). It follows by Theorem \ref{theo Germany}, by (i) and since $G$ is prime.  \qed  \\

{\bf Remark 1.} According to Lemma \ref{lemm Y_i} (ii), throughout the remaining part of the paper let us denote as $y_i$ the possible vertex of $Y_i$ for $i = 1,\ldots,5$.

\begin{lemm}\label{lemm Z}

$Z_{i-1,i+1}$ has a join to $\{y_{i+2},y_{i+3}\}$, for $i = 1,\ldots,5$.

\end{lemm}

{\bf Proof.} Let $z \in Z_{i-1,i+1}$.

First assume that $y_i$ does exist. Then by Lemma \ref{lemm 1 Germany}, $y_i$ is adjacent to $y_{i+2},y_{i+3}$. Then $z$ is adjacent to $y_{i+2}$, otherwise $v_{i+4},z,v_{i+1},v_{i+2},y_{i+2},y_i$ induce a $P_6$. Similarly by symmetry $z$ is adjacent to $y_{i+3}$.

Then assume that $y_i$ does not exist and that there is $z^0 \in Z_{i,i+2} \cup Z_{i,i+3}$ nonadjacent to $z$, say $z^0 \in Z_{i,i+2}$ without loss of generality by symmetry. Then $y_{i+2}$ is adjacent to $z$, otherwise $y_{i+2},v_{i+2},z^0,v_i,v_{i+4},z$ induce a $P_6$. Then let us consider $y_{i+3}$. Assume by contradiction that $z$ is nonadjacent to $y_{i+3}$.
%Notice that $y_{i+3}$ is nonadjacent to $z^0$, otherwise $v_{i+3},y_{i+3},z^0,v_i,v_{i+1},z$ induce a $P_6$. Then $y_{i+1}$ does not exist, otherwise $y_{i+1},y_{i+3},v_{i+3},v_{i+4},v_i,z^0$ induce a $P_6$.
By Lemma \ref{lemm G_0 Germany}, $z$ is adjacent to each vertex of $Z_{i,i+2} \setminus \{z^0\}$. Then to avoid that $\{v_i,z\}$ and $\{z^0\}$ form a bi-module of $G$, either $z$ is adjacent to a vertex $z' \in Z_{i,i+3}$ nonadjacent to $z^0$, that is $z,v_{i+1},v_i,z',v_{i+3},y_{i+3}$ induce a $P_6$, or $z$ is adjacent to a vertex $x$ of $X_T$, that is $x,z,v_{i+1},v_{i+2},v_{i+3},y_{i+3}$ induce a $P_6$, a contradiction.

Finally assume that $y_i$ does not exist and that $z$ dominates $Z_{i,i+2} \cup Z_{i,i+3}$. Then to avoid that $\{v_i,z\}$ forms a module of $G$, $z$ is adjacent to at least one vertex $q$ from $X_T \cup \{y_{i+2},y_{i+3}\}$, and to avoid that $\{v_i,z\}$ and $\{q\}$ form a bi-module of $G$, $z$ is adjacent to at least two vertices from $X_T \cup \{y_{i+2},y_{i+3}\}$ (recall that $z$ is adjacent to at most one vertex of $X_T$). Then $z$ is adjacent to at least one vertex from $\{y_{i+2},y_{i+3}\}$, say $y_{i+2}$ without loss of generality by symmetry. Moreover, if $y_{i+3}$ does exist, then $z$ is adjacent to $y_{i+3}$ as well, otherwise $x \in X_T$, $z,v_{i+1},v_{i+2},v_{i+3},y_{i+3}$ induce a $P_6$.      \qed \\

%Notice that if $z$ is adjacent to $y_{i+2}$ then $z$ is adjacent to $y_{i+3}$ as well (otherwise $v_{i+1},z,$)
%First assume that $z \in Z^1_{i-1,i+1}$, then: to avoid that $\{v_i,z\}$ and $\{y_i\}$ form a bi-module of $G$, $z$ is adjacent to at least one vertex from $X_T \cup \{y_{i+2},y_{i+3}\}$ (recall that by definition of $X_T$, $z$ is adjacent to at most one vertex of $X_T$); in particular, if $z$ is adjacent to $x \in X_T$, then

Let us consider $X_T$. Recall that each vertex of $Z$ is adjacent to at most one vertex of $X_T$.

\begin{lemm}\label{lemm X_T}

If $x \in X_T$ is adjacent to $z \in Z^1_{i-1,i+1}$, then one of the following cases occurs:

\begin{enumerate}
\item[$(i)$] $X_T = \{x\}$ and $Z = \{z,\bar{z}\}$ with $\bar{z} \in Z_{i+1,i+3}$;
\item[$(ii)$] $X_T = \{x\}$ and $Z = \{z,\bar{z},\tilde{z}\}$ with $\bar{z} \in Z_{i+1,i+3}$, $\tilde{z} \in Z_{i+2,i+4}$, $z,\bar{z},\tilde{z}$ mutually nonadjacent, and $x$ adjacent to each of them.
\end{enumerate}

%\begin{enumerate}
%\item[$(i)$] $Z^1_{i-1,i+1} = \{z\}$;
%\item[$(ii)$] $x$ is adjacent to no vertex of $Z^0_{i-1,i+1}$;
%\item[$(iii)$] $\{y_{i+2},y_{i+3}\} \neq \emptyset$, and $z$ is adjacent to $y_{i+2}$ (if) and $y_{i+3}$ (if);
%\item[$(iii)$] $x$ is adjacent to $\bar{z} \in Z_{i+1,i+3} \cup Z_{i+2,i+4}$; in particular: (a) if $\bar{z} \in Z^1$, then $Z = \{z,\bar{z}\}$, (b) if $\bar{z} \in Z^0$, then $Z = \{z,\bar{z},\tilde{z}\}$ with $\bar{z}$ and $\tilde{z}$ respectively in $Z^0_{i+1,i+3}$ and in $Z^0_{i+2,i+4}$ (and $x$ adjacent to both);
%\item[$(iv)$] $X_T$ has a co-join to $Z_{i,i+2} \cup Z_{i,i+3}$, and to $Z^0_{i-1,i+1}$.  \qed
%\end{enumerate}

\end{lemm}

{\bf Proof.} The proof is given by the following claims.

{\bf Claim 1} {\em $x$ is adjacent to a vertex $\bar{z} \in Z_{i+1,i+3} \cup Z_{i+2,i+4}$, say $\bar{z} \in Z_{i+1,i+3}$}

{\bf proof.} Let us observe that $x$ is adjacent to no vertex $a \in Z^1_{i-1,i+1} \setminus \{z\}$, otherwise $\{a,z\}$ forms a module of $G$. Then since $z$ can not have degree 1 (by Theorem \ref{theo clique cutset}),
%(by Lemma \ref{lemma 3 Mahjoub}),
$z \in Z^1_{i-1,i+1}$ and $G$ is triangle-free, the claim follows. \qed

{\bf Claim 2} {\em $Z^1_{i-1,i+1} = \{z\}$}

{\bf proof.} Assume to the contrary that there is $z' \in Z^1_{i-1,i+1}$ different to $z$.

If $\{y_{i+2},y_{i+3}\} = \emptyset$, then to avoid that either $\{v_i,z'\}$ forms a module of $G$ or $\{v_i,z'\}$ and $\{y_i\}$ form a bi-module of $G$, $z'$ is adjacent to $x$; then $\{z,z'\}$ forms a module of $G$, a contradiction.

Then assume that $\{y_{i+2},y_{i+3}\} \neq \emptyset$. By Lemma \ref{lemm Z} $z$ and $z'$ are adjacent to $y_{i+2},y_{i+3}$. Then to avoid that $\{z,z'\}$ is a module of $G$, $z'$ is nonadjacent to $x$. Notice that $y_{i+2}$ does not exist, otherwise by Lemma \ref{lemm Y_i} $x,\bar{z},v_{i+3},v_{i+4},z',y_{i+2}$ induce a $P_6$. Then there exists $y_i$, otherwise $\{v_i,z'\}$ and $\{y_{i+3}\}$ form a bi-module of $G$. Then $x,\bar{z},y_i,y_{i+3},z',v_{i+4}$ induce a $P_6$, a contradiction.  \qed

{\bf Claim 3} {\em $x$ is adjacent to no vertex of $Z^0_{i-1,i+1}$}

{\bf proof.} Assume to the contrary that $x$ is adjacent to $z^0 \in Z^0_{i-1,i+1}$. To avoid that $\{z,z^0\}$ forms a module of $G$, there is $q \in Z_{i,i+2} \cup Z_{i,i+3}$, say without loss of generality $q \in Z_{i,i+3}$, adjacent to $z$ and nonadjacent to $z^0$. Then $x$ is adjacent to $q$, otherwise $z^0,x,z,q,v_{i+3},v_{i+2}$ induce a $P_6$. Then $x,q,z$ induce a triangle, a contradiction.   \qed

%Proof of (iii). By (i) $Z^1_{i-1,i+1} = \{z\}$. Then to avoid that $\{v_i,z\}$ and $\{x\}$ form a bi-module of $G$, $z$ is adjacent to at least two vertices not in $N(v_i)$, i.e., to at least one vertex from $\{y_{i+2},y_{i+3}\}$; in particular to avoid a $P_6$ (involving $x$), $z$ is adjacent to both those vertices (if both).

{\bf Claim 4} {\em If $\bar{z} \in Z^1_{i+1,i+3}$, then case (i) occurs}.

{\bf proof.} Let us show that $Z_0 = \emptyset$. Assume by contradiction that there is $z^0 \in Z_0$ and consider the following occurrences which are exhaustive by symmetry. If $z^0 \in Z^0_{i,i+2}$, then by definition of $Z^0$ there is $q \in Z^0_{i-1,i+1}$ (without loss of generality, by symmetry) nonadjacent to $z^0$; then $q,v_{i-1},v_i,z^0,\bar{z},x$ induce a $P_6$, a contradiction. If $z^0 \in Z^0_{i-1,i+1}$, then by definition of $Z^0$ and by the previous fact there is $q' \in Z^0_{i,i+3}$ nonadjacent to $z^0$: then $y_{i+2}$ does not exist, otherwise by Lemma \ref{lemm Z} $q',z,y_{i+2}$ induce a triangle; also $y_i$ does not exist, otherwise $x,\bar{z},y_i,v_i,v_{i+4},z^0$ induce a $P_6$; then to avoid that $\{v_i,z\}$ and $\{x\}$ form a bi-module of $G$, $y_{i+3}$ does exist (adjacent to $z$); also to avoid that $\{v_{i+2},\bar{z}\}$ and $\{x\}$ form a bi-module of $G$, $y_{i+4}$ does exist (adjacent to $\bar{z}$); then $z^0,y_{i+3},z,x,\bar{z},y_{i+4}$ induce a $P_6$, a contradiction. If $z^0 \in Z^0_{i,i+3}$, then by definition of $Z_0$ and by the previous fact there is $q'' \in Z^0_{i+2,i+4}$ nonadjacent to $z^0$; then $v_i,z^0,z,x,\bar{z},q''$ induce a $P_6$. Then the assertion follows.

Let us show that $Z_1 = \{z,\bar{z}\}$. By Claim 2 and its symmetric version, $Z^1_{i-1,i+1} = \{z\}$ and $Z^1_{i+1,i+3} = \{\bar{z}\}$. Consider $Z^1_{i,i+2}$. Assume by contradiction that there is $q \in Z^1_{i,i+2}$. If $y_{i+1}$ does exist, then: if $q$ is adjacent to a vertex $x'$ of $X_T$ ($x' \neq x$ to avoid a triangle), then by the symmetric version of Claim 1 and since $Z_0 = \emptyset$, there is $q' \in Z^1_{i+2,i+4}$ (w.l.o.g.) adjacent to $x'$ (and to $y_{i+1}$, by Lemma \ref{lemm Z}), and consequently $x,z,q,x',q',y_{i+1}$ induce a $P_6$; if $q$ is adjacent to no vertex of $X_T$, then to avoid a bi-module of $G$, $q$ is adjacent either to $y_{i+3}$ (thus $q,y_{i+3},z$ induce a triangle) or to $y_{i+4}$ (thus $q,y_{i+4},\bar{z}$ induce a triangle). If $y_{i+1}$ does not exist, then to avoid a bi-module of $G$, $q$ is adjacent either to $y_{i+3}$ or to $y_{i+4}$, a contradiction similar to the previous sentence. Then $Z^1_{i,i+2} = \emptyset$. Consider $Z^1_{i,i+3}$. Assume by contradiction that there is $p \in Z^1_{i,i+3}$. Then $p$ is adjacent to no vertex $x'$ of $X_T$ ($x' \neq x$ to avoid a triangle), otherwise $x',p,v_i,v_{i+1},\bar{z},x$ induce a $P_6$. Also, $y_{i+4}$ does not exist, otherwise $v_i,p,z,x,\bar{z},y_{i+4}$ induce a $P_6$. Then to avoid a bi-module of $G$, $p$ is adjacent to $y_{i+1},y_{i+2}$ (i.e., they exist), and consequently by Lemma \ref{lemm Z} $p,y_{i+2},z$ induce a triangle, a contradiction. Then $Z^1_{i,i+3} = \emptyset$. Similarly, by symmetry, $Z^1_{i+2,i+4} = \emptyset$. Then the assertion follows.     \qed

{\bf Claim 5} {\em If $\bar{z} \in Z^0_{i+1,i+3}$, then case (ii) occurs}.

{\bf proof.} Let us show that $Z_0 = \{\bar{z},\tilde{z}\}$, where $\tilde{z}$ is a vertex of $Z^0_{i+2,i+4}$. First let us observe that: $Z^0_{i-1,i+1} = \emptyset$, otherwise by Lemma \ref{lemm G_0 Germany} $x$ should be adjacent to a vertex of $Z^0_{i-1,i+1} = \emptyset$, a contradiction to Claim 3; $Z^0_{i,i+2} = Z^0_{i,i+3} = \emptyset$, otherwise by Lemma \ref{lemm G_0 Germany} $x$ should be adjacent to a vertex of such sets, but then a triangle arises since $z \in Z^1_{i-1,i+1}$. Then by definition of $Z_0$ and by Lemma \ref{lemm G_0 Germany} there exists $\tilde{z} \in Z^0_{i+2,i+4}$, with $\tilde{z}$ adjacent to $x$ by Lemma \ref{lemm G_0 Germany}. If $y_i$ does exist, then $(Z^0_{i+1,i+3} \setminus \{\bar{z}\}) \cup (Z^0_{i+2,i+4} \setminus \{\tilde{z}\}) = \emptyset$, otherwise since by Lemma \ref{lemm G_0 Germany} $\tilde{z}$ ($\bar{z}$) is adjacent to each vertex of $Z^0_{i+1,i+3} \setminus \{\bar{z}\}$ (of $Z^0_{i+2,i+4} \setminus \{\tilde{z}\}$), Lemma \ref{lemm Z} imply that $G$ has a triangle involving $\tilde{z}$ or $\bar{z}$ respectively. If $y_i$ does not exist, then to avoid a bi-module of $G$ involving $z$, at least one from $\{y_{i+2},y_{i+3}\}$ does exist, say $y_{i+2}$ without loss of generality; then $Z^0_{i+1,i+3} \setminus \{\bar{z}\} = \emptyset$ (otherwise $y_{i+2},z,x,\tilde{z}$, a vertex of $Z^0_{i+1,i+3}$ and $v_{i+3}$ induce a $P_6$), and consequently $Z^0_{i+2,i+4} \setminus \{\tilde{z}\} = \emptyset$ (by definition of $Z_0$ and by Lemma \ref{lemm G_0 Germany}). Then the assertion follows.

Let us show that $Z_1 = \{z\}$. By Claim 2, $Z^1_{i-1,i+1} = \{z\}$. Notice that $Z^1_{i,i+2} = \emptyset$: in fact if there is $q \in Z^1_{i,i+2}$, then $q$ is nonadjacent to $x$ (to avoid a triangle) and in general to no vertex $x' \in X_T$ (to avoid that $x',q,\bar{z},x,\tilde{z},v_{i+4}$ induce a $P_6$); then to avoid a bi-module of $G$, $q$ is adjacent either to $y_{i+3}$ (and thus $q,y_{i+3},z$ form a triangle) or to $y_{i+4}$ (and thus $q,y_{i+3},\bar{z}$ form a triangle). Similarly by symmetry, $Z^1_{i,i+2} = \emptyset$. Furthermore, $Z^1_{i+1,i+3} = \emptyset$: in fact if there is $p \in Z^1_{i,i+2}$, then $p$ is adjacent to $x$ (to avoid a triangle) and in general to no vertex $x' \in X_T$ (to avoid that $x',p,v_{i+3},\bar{z},x,z$ induce a $P_6$); then to avoid a bi-module of $G$, $p$ is adjacent either to $y_i$ (and thus $p,y_i,\tilde{z}$ form a triangle) or to $y_{i+4}$ (this means, by the previous facts, that $y_{i+2}$ does exist to avoid that $\{v_{i+2},p\}$ and $\{y_{i+4}\}$ form a bi-module, and consequently that $x,z,y_{i+2},y_{i+4},p,v_{i+3}$ induce a $P_6$). Similarly, by symmetry, $Z^0_{i+2,i+4} = \emptyset$. Then the assertion follows.          \qed

Then the lemma is proved. \qed  \\

Let us consider $Z$ and subgraph $G_0$.

\begin{lemm}\label{lemm Z^1}
The following facts hold for $i = 1,\ldots,5$:

\begin{enumerate}
\item[$(i)$] $|Z^1_{i-1,i+1}| \leq 1$;
\item[$(ii)$] if there is an edge between $Z_{i+1,i+3}$ and $Z_{i+2,i+4}$, then $Y_{i} = Z^1_{i-1,i+1} = \emptyset$;
\item[$(iii)$] if $Z^0_{i+1,i+3} \neq \emptyset$ and $Z^0_{i+2,i+4} \neq \emptyset$, then $Z^1_{i,i+2} = Z^1_{i,i+3} = \emptyset$;
\item[$(iv)$] if $Z^0_{i+1,i+3} \neq \emptyset$ and $Z^0_{i+2,i+4} \neq \emptyset$ and $Z^0_{i,i+3} \neq \emptyset$, then $Z_1 = \emptyset$.
\end{enumerate}
\end{lemm}

{\bf Proof.} Proof of (i). It follows by Theorem \ref{theo Germany}, Lemmas \ref{lemm Z} and \ref{lemm X_T}, and since $G$ is prime.

Proof of (ii). Let $a \in Z_{i+1,i+3}$ be adjacent to $b \in Z_{i+2,i+4}$. By Lemma \ref{lemm Z} and since $G$ is triangle-free, one has $Y_{i} = \emptyset$. Let us consider $Z^1_{i-1,i+1}$. Assume by contradiction that there is $z \in Z^1_{i-1,i+1}$. By Lemma \ref{lemm X_T}, $z$ is adjacent to no vertex of $X_T$. Then, since $Y_{i} = \emptyset$, to avoid that either $\{v_{i},z\}$ form either a module of $G$ or $\{v_{i},z\}$ and one vertex from $\{y_{i+2},y_{i+3}\}$ form either a module of $G$, both $y_{i+2}$ and $y_{i+3}$ do exist. Then $y_{i+2},z,y_{i+3},v_{i+3},a,b$ induce a $P_6$, a contradiction.

Proof of (iii). First let us observe that there exist $a \in Z^0_{i+1,i+3}$ and $b \in Z^0_{i+2,i+4}$ mutually nonadjacent: in fact otherwise, by definition of $Z^0$, there are $a' \in Z_{i,i+2}$ nonadjacent to $a$, and $b' \in Z_{i,i+3}$ nonadjacent to $b$, with $a',v_{i},b',v_{i+3},a,b$ induce a $P_6$.

Let us consider $Z^1_{i,i+2}$. Assume by contradiction that there is $z \in Z^1_{i,i+2}$. Then: $z$ is adjacent to no vertex of $X_T$, by Lemma \ref{lemm X_T}; $y_{i+4}$ does not exist, otherwise by Lemma \ref{lemm Z} $z,a,y_{i+4}$ induce a triangle. If $y_{i+1}$ does not exist, then either $\{v_{i+1},z\}$ forms a module of $G$ or $\{v_{i+1},z\}$ and $\{y_{i+3}\}$ form a bi-module of $G$. If $y_{i+1}$ does exist, then $y_{i+3}$ does exist too, otherwise $\{v_{i+1},z\}$ and $\{y_{i+1}\}$ form a bi-module of $G$: then $a,z,y_{i+3},y_{i+1},b,v_{i+4}$ induce a $P_6$, a contradiction. Then $Z^1_{i,i+2} = \emptyset$. Similarly by symmetry $Z^1_{i,i+3} = \emptyset$.

Proof of (iv). By (iii), $Z_1 \setminus Z^1_{i+2,i+4} = \emptyset$. Assume by contradiction that there is $z \in Z^1_{i+2,i+4}$. Then to avoid that either $\{v_{i+3},z\}$ forms a module of $G$ or $\{v_{i+3},z\}$ and $\{y_{i+3}\}$ form a bi-module of $G$, there exists at least one vertex from $\{y_i,y_{i+1}\}$: then a triangle arises by Lemma \ref{lemm Z}. \qed  \\

{\bf Remark 2.} According to Lemma \ref{lemm Z^1} (i), throughout the remaining part of the paper let us denote as $z_{i-1,i+1}$ the possible vertex of $Z^1_{i-1,i+1}$ for $i = 1,\ldots,5$.

\begin{lemm}\label{lemm case(iii)}

Case (iii) of Lemma \ref{lemm G_0 Germany} can not occur.

\end{lemm}

{\bf Proof.} Let $Z^0_{i+1,i+3} \cup Z^0_{i+2,i+4}$ induce the co-matched bipartite graph forming $G[Z_0]$. In particular, one can assume that there is at least one edge from $Z^0_{i+1,i+3}$ to $Z^0_{i+2,i+4}$ (otherwise case (ii) of Lemma \ref{lemm G_0 Germany} occurs). Then by Lemma \ref{lemm Z^1} (ii), $y_i$ does not exist, and by Lemma \ref{lemm Z^1} (iii), $Z_1 \setminus (Z^1_{i+1,i+3} \cup Z^1_{i+2,i+4}) = \emptyset$. Then in $Z_1$ there are at most $z_{i+1,i+3}$ and $z_{i+2,i+3}$. Then: if $X_0 = \emptyset$, then $Z^0_{i+1,i+3}$ and $Z^0_{i+2,i+4}$ form a bi-module of $G$, which is not possible by Lemma \ref{lemm bi-module}; if $X_0 \neq \emptyset$, then one can check that $G$ is ferry, which is not possible by Lemma \ref{F_m}. \qed

\begin{lemm}\label{lemm case(iv)}

Case (iv) of Lemma \ref{lemm G_0 Germany} can not occur.

\end{lemm}

{\bf Proof.} Let $Z^0_{i+1,i+3} \cup Z^0_{i+2,i+4}$ induce one of the two co-matched bipartite graph forming $G[Z_0]$. In particular, one can assume that there is at least one edge from $Z^0_{i+1,i+3}$ to $Z^0_{i+2,i+4}$ (otherwise case (ii) of Lemma \ref{lemm G_0 Germany} occurs). Then, since $X_0 = \emptyset$, one can prove the lemma by applying the argument of Lemma \ref{lemm case(iii)}.    \qed  \\

%By symmetry it is sufficient to consider just one of such co-matched bipartite graphs forming $G[Z_0]$, say that induced by $Z^0_{i+1,i+3} \cup Z^0_{i+2,i+4}$. By Proposition \ref{prop 2 unbounded} the only vertex which can distinguish respectively $Z^0_{i+1,i+3}$ and $Z^0_{i+2,i+4}$ is $y_i$. If case (i) of Proposition \ref{prop 1 unbounded} occurs, then $Z^0_{i+1,i+3}$ and $Z^0_{i+2,i+4}$ form a bi-module of $G$, i.e. they induce an edge by Lemma \ref{lemm bi-module}, a contradiction. If case (ii) of Proposition \ref{prop 1 unbounded} occurs, then let $z',z''$ with $z' \in Z^0_{i+1,i+3}$ and $z'' \in Z^0_{i+2,i+4}$ according to the corresponding statement; then $|Z^0_{i+1,i+3} \setminus \{z'\}| \leq 1$ and $|Z^0_{i+2,i+4} \setminus \{z''\}| \leq 1$ otherwise they form a bi-module of $G$ (i.e. they would induce an edge by Lemma \ref{lemm bi-module}, a contradiction). Then the lemma follows, by considering also the possible vertex of $X_0$ adjacent to $z',z''$.        \qed

Then let us summarize the above results by the following theorem.

\begin{theo}\label{theo structure}
Every prime facet-inducing ($P_6$,triangle)-free graph is a subgraph of one the graphs $H_1,H_2,H_3$ drawn respectively in Figures 2, 3, 4.
\end{theo}

{\bf Proof.} Let $G$ be a (non-empty) prime facet-inducing ($P_6$,triangle)-free graph. If $G$ is bipartite, then $G$ is perfect, i.e., $G = K_2$. Then assume that $G$ is not bipartite. By Lemmas \ref{lemm case(iii)} and \ref{lemm case(iv)}, only cases (i) and (ii) of Lemma \ref{lemm G_0 Germany} may occur. By the other results of this section, $G$ has at most 21 vertices. In fact, one has $|Y_i| \leq 1, |Z^1_i| \leq 1, |Z^0_i| \leq 1$ for $i = 1,\ldots,5$; furthermore, $X'_i = \emptyset$, and $|X_T| \leq 1$ (since: if $x \in X_T$ is adjacent to a vertex of $Z_1$, then Lemma \ref{lemm X_T} holds; otherwise, Lemma \ref{lemm G_0 Germany} (i)-(ii) holds).

Then let us focus on the following cases.

{\bf Case 1} Case (i) of Lemma \ref{lemm G_0 Germany} occurs.

This means that $Z_0 = \emptyset$. Then $G$ is a subgraph of the graph $H_1$ where the vertices of $Z$ are gray and the vertex of $X_T$ is the big central one.

\begin{figure}
\centering
\includegraphics[width=\textwidth]{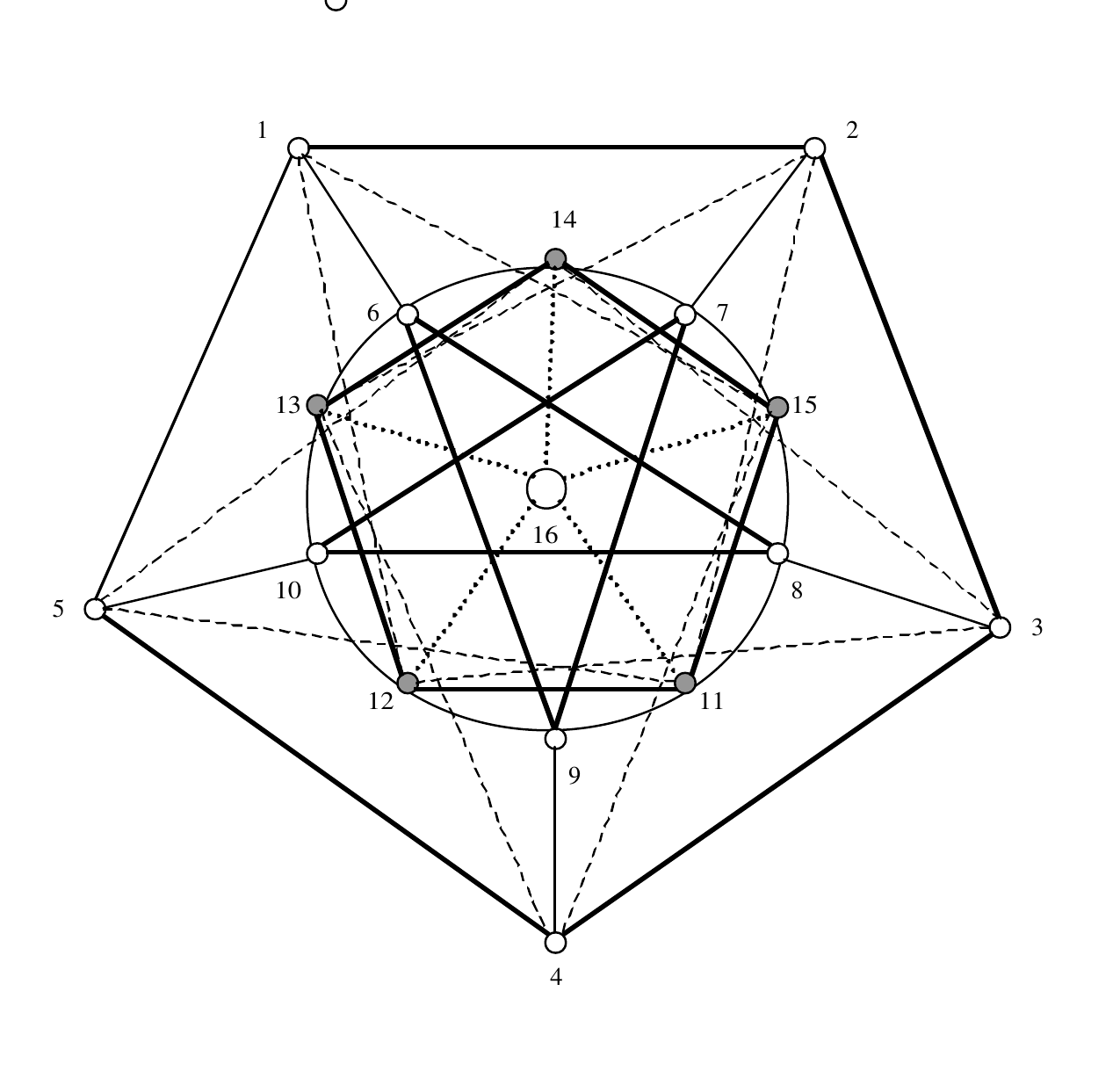}
\caption{The graph $H_1$}
\end{figure}

{\bf Case 2} Case (ii) of Lemma \ref{lemm G_0 Germany} occurs.

{\bf Subcase 2.1} $Z_1 = \emptyset$.

Then $G$ is a subgraph of the graph $H_2$ where the vertices of $Z$ are gray and the vertex of $X_T$ is the big central one.

\begin{figure}
\centering
\includegraphics[width=\textwidth]{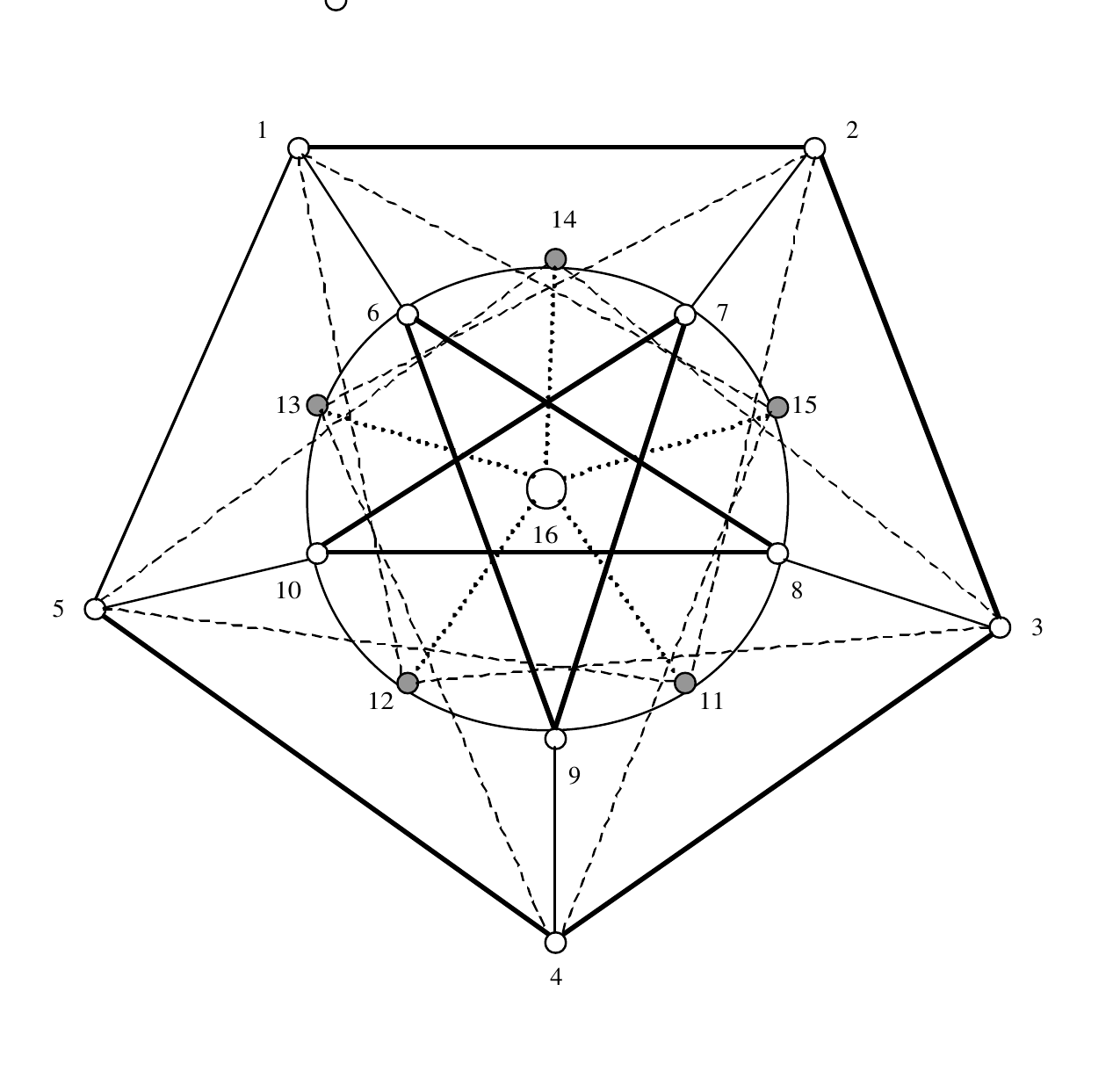}
\caption{The graph $H_2$}
\end{figure}

{\bf Subcase 2.2} $Z_1 \neq \emptyset$ and the vertex of $X_T$ is adjacent to a vertex of $Z_1$.

Then by Lemma \ref{lemm X_T}, $G$ is a subgraph of the graph $H_2$ where the vertices of $Z$ are gray and the vertex of $X_T$ is the big central one.

{\bf Subcase 2.3} $Z_1 \neq \emptyset$ and the vertex of $X_T$ is adjacent to no vertex of $Z_1$.

Then by Lemma \ref{lemm Z} and by definition of $Z_0$, there are exactly two consecutive vertices of $Z_0$, say $z^0_{2,4},z^0_{3,5}$, plus possibly: $y_2,y_3,y_4,y_5$ (by Lemma \ref{lemm Z} $y_1$ does not exist), $z_{2,4},z_{3,5},z_{2,5}$, and a vertex $x \in X_0$ adjacent only to $z^0_{2,4},z^0_{3,5}$. Then $G$ is a subgraph of the graph $H_3$ where the vertices of $Z$ are gray and the vertex of $X_T$ is the big central one. \qed \\

Let us observe that $H_1$ is not ($P_6$,triangle)-free, while $H_2$ and $H_3$ are ($P_6$,triangle)-free.

\begin{figure}
\centering
\includegraphics[width=\textwidth]{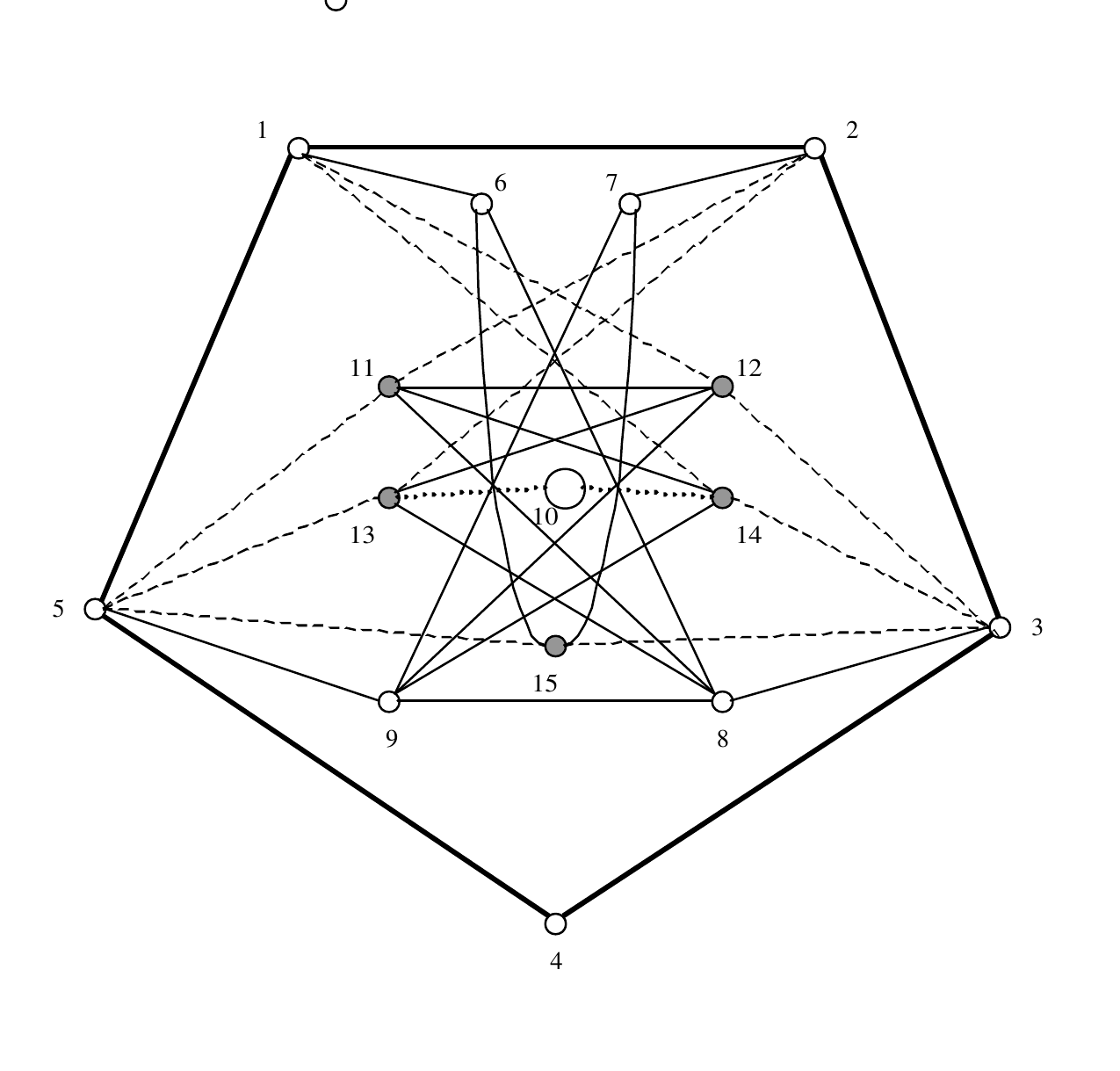}
\caption{The graph $H_3$}
\end{figure}

\section{The stable set polytope of ($P_6$,triangle)-free graphs}

In this section let us describe the stable set polytope of ($P_6$,triangle)-free graphs (and more generally of ($P_6$,paw)-free graphs).  \\

{\bf Remark 3.} The stable set polytope of specific graphs has been computed by an adapted version of the software PORTA (available on line www.zib.de/Optimization/Software/Porta/), which Prof. Caterina De Simone (IASI-CNR, Rome) kindly sent me.

\begin{theo}\label{theo final}
Let ${\cal X}$ be the class of ($P_6$,triangle)-free graphs. Then ${\cal F_P(X)} = \{K_2,C_5,G_1,\ldots,G_{24}\}$, where $G_1$ be the graph drawn in Figure 5, and referring to the statement of Theorem \ref{theo structure}: $G_2 = H_2$, $G_3 = H_2 - \{1\}$, $G_4 = H_2 - \{1,2\}$, $G_5 = H_2 - \{1,2,3\}$, $G_6 = H_2 - \{1,2,4\}$, $G_7 = H_2 - \{1,2,3,4\}$, $G_8 = H_2 - \{1,2,3,12\}$, $G_9 = H_2 - \{1,2,3,13\}$, $G_{10} = H_2 - \{1,2,3,4,5\}$, $G_{11} = H_2 - \{1,2,3,4,12\}$, $G_{12} = H_2 - \{1,2,3,4,5,11\}$, $G_{13} = H_2 - \{1,2,3,4,5,11,14\}$, $G_{14} = H_3$, $G_{15} = H_3 - \{4\}$, $G_{16} = H_3 - \{15\}$, $G_{17} = H_3 - \{4,12\}$, $G_{18} = H_3 - \{4,11,12\}$, $G_{19} = H_3 - \{4,9,12\}$, $G_{20} = H_3 - \{4,5,11,12\}$, $G_{21} = H_3 - \{4,10,12,13\}$, $G_{22} = H_3 - \{4,7,9,11,12\}$, $G_{23} = H_3 - \{4,10,11,12,13\}$, $G_{24} = H_3 - \{4,10,11,12,13,14\}$.

\end{theo}

{\bf Proof.} By Theorem \ref{theo structure}, the elements of ${\cal F_P(X)}$ are subgraphs of $H_1,H_2,H_3$. Then they can be detected by computing the stable set polytope of graphs $H_1,H_2,H_3$, i.e., one has to detect the prime facet-inducing subgraphs of those three graphs. Then let us refer to Remark 3.

STAB($H_1$) provides the following facet-inducing subgraphs up to isomorphism: $K_2,C_5,G_1$ (they are the only ($P_6$,triangle)-free ones).

STAB($H_2$) provides the following facet-inducing subgraphs up to isomorphism: $K_2,C_5,G_1,\ldots,G_{13}$.

STAB($H_3$) provides the following facet-inducing subgraphs up to isomorphism: $K_2,C_5,G_1,G_{14},\ldots,G_{24}$. \qed \\

\begin{figure}
\centering
\includegraphics[width=\textwidth]{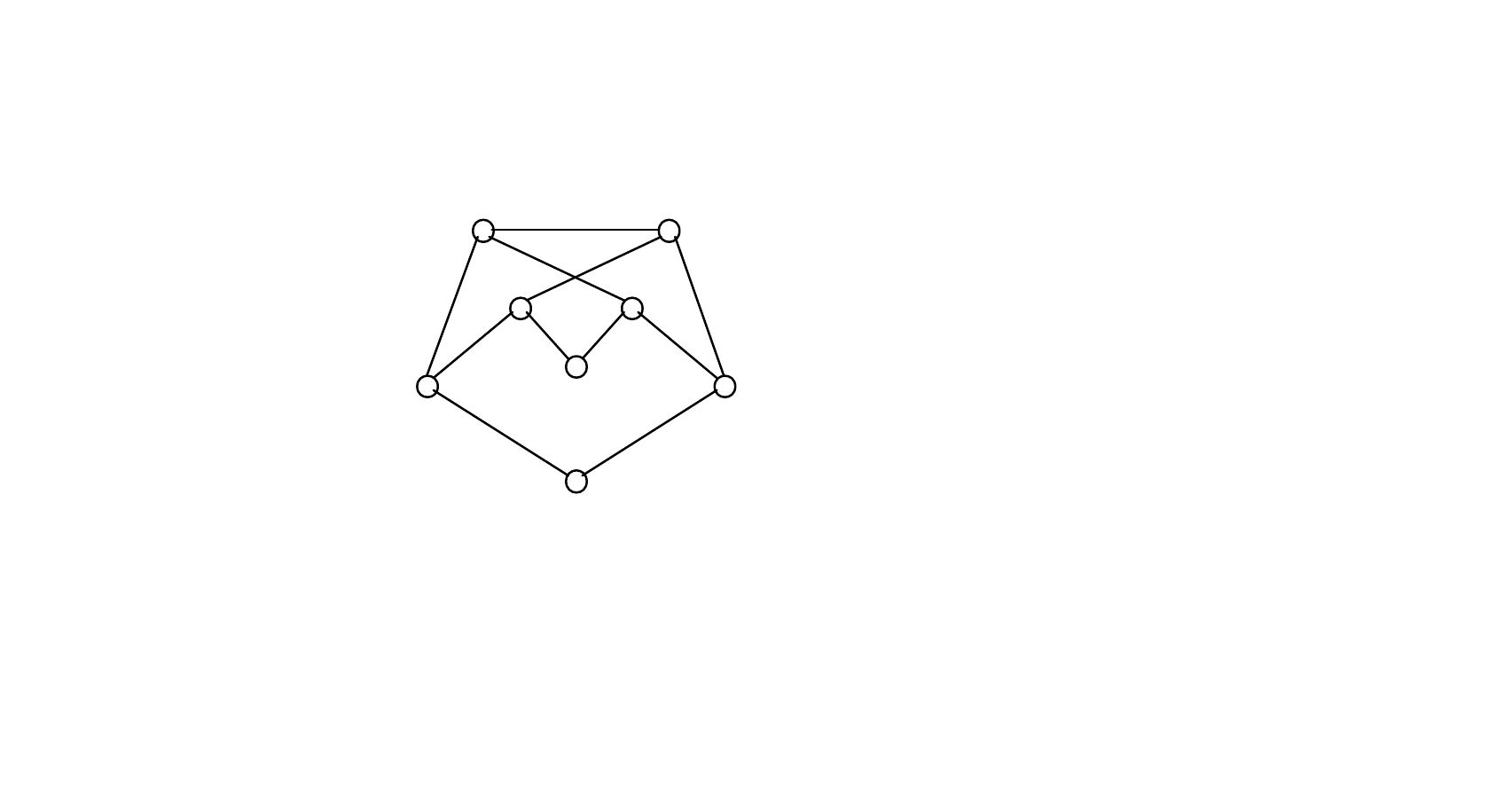}
\caption{The graph $G_1$}
\end{figure}

Then by Proposition \ref{prop trivial}, Observation \ref{obse triangle-free}, and Theorem \ref{theo final} one obtains:

\begin{theo}\label{theo final final}
Let $G$ be a ($P_6$,triangle)-free graph. Then a (minimal) linear system
of STAB($G$) is given by:
\begin{itemize}
\item[($a$)] $-x_i\leq 0$ for every node $i$ of $G$;
\item[($b$)] $\{\Phi (H) : H \in \{K_2,C_5,G_1,\ldots,G_{24}\}\}$, where graphs $G_1,\ldots,G_{24}$ are defined in the statement of Theorem \ref{theo final}.     \qed
\end{itemize}
\end{theo}

Let us compare this result with that concerning the class, say ${\cal X'}$, of ($P_5$,triangle)-free graphs:
as one can check from \cite{DeSMos}, or from Theorem \ref{theo final} by ignoring those graphs containing a $P_5$, one has ${\cal F_P(X')} = \{K_2,C_5\}$; in particular, according to Observation \ref{obse triangle-free}, ($P_5$,triangle)-free graphs are $t$-perfect \cite{GroLovSch1998}.

Then let us consider ($P_6$,paw)-free graphs. Let us report the following result of Olariu \cite{Olariu1998}.

\begin{theo}[\cite{Olariu1998}]\label{Olariu}
Every connected paw-free graph is either triangle-free or complete multipartite (that is it admits a partition into stable sets which have mutually a join). \qed
\end{theo}

By combining Theorem \ref{Olariu} with the above one obtains the following corollaries.

\begin{coro}\label{coro final}
Let ${\cal Y}$ be the class of ($P_6$,paw)-free graphs. Then ${\cal F_P(Y)} = \{K_2,C_5,G_1,\ldots,G_{24}\}$, where graphs $G_1,\ldots,G_{24}$ are defined in the statement of Theorem \ref{theo final}. \qed
\end{coro}

{\bf Proof.} Let $G$ be a prime facet-inducing ($P_6$,claw)-free graph. Since every complete multipartite graph is perfect, and since every perfect prime facet-inducing graph is a $K_2$, Theorem \ref{Olariu} implies that $G$ is triangle-free. Then the corollary follows by Theorem \ref{theo final}. \qed  \\

Let $F$ be a subgraph of a graph $G$. Let $C(F,G)$ the family of subgraphs $H$ of $G$ such that: (i) $H$ contains $F$, and (ii) if $H$ properly contains $F$, then $H - F$ is a clique and has a join to $F$. In particular $F \in C(F,G)$. An element $H$ of $C(F,G)$ is $maximal$ if no element of $C(F,G)$ properly contains $H$. Let $C^*(F,G)$ be the family of maximal elements of $C(F,G)$. For instance, $C^*(K_2,G)$ is the family of maximal cliques of $G$.

\begin{coro}\label{coro final final}
Let $G$ be a ($P_6$,paw)-free graph. Then a (minimal) linear system
of STAB($G$) is given by:
\begin{itemize}
\item[($a$)] $-x_i\leq 0$ for every node $i$ of $G$;
\item[($b$)] $\{\Phi (H) : H \in C^*(K_2,G) \cup C^*(C_5,G) \cup C^*(G_1,G) \cup \ldots \cup C^*(G_{24},G)\}$, where graphs $G_1,\ldots,G_{24}$ are defined in the statement of Theorem \ref{theo final}.         \qed
\end{itemize}
\end{coro}

{\bf Proof.} Let ${\cal Y}$ be the class of ($P_6$,paw)-free graphs. By Corollary \ref{coro substitution}, ${\cal F(Y)} = {\cal S(F_P(Y))} \cap {\cal Y}$. The set ${\cal S(F_P(Y))} \cap {\cal Y}$ depends on ${\cal F_P(Y)}$, which is given by Corollary \ref{coro final}. Then any graph in ${\cal F_P(Y)}$ can be (repeatedly) substituted for vertices of just one graph in ${\cal F_P(Y)}$), namely of $K_2$, since otherwise a paw arises. That is, ${\cal F(Y)}$ is given by $C(K_2,G) \cup C(C_5,G) \cup C(G_1,G) \cup \ldots \cup C(G_{24},G)$. Then for a minimal description of STAB($G$) one may just consider $C^*(K_2,G) \cup C^*(C_5,G) \cup C^*(G_1,G) \cup \ldots \cup C^*(G_{24},G)$. Then the corollary follows by Proposition \ref{prop trivial}. \qed  \\

The Maximum (Weight) Stable Set Problem and the Maximum (Weight) Clique Problem for ($P_6$,triangle)-free graphs can be solved in polynomial time (i.e., $O(n^2)$ time) \cite{BraKleMah2005}. That can be extended to ($P_6$,paw)-free graphs by Theorem \ref{Olariu}. Then the following fact concerns the separation problem $-$ see e.g. \cite{GroLovSch1998,Pulleyblank1989,Schrijver1995,Schrijver2003}

\begin{theo}
The separation problem for STAB($G$) by facets can be solved in polynomial time when $G$ belongs to the class of ($P_6$,triangle)-free graphs or more generally of ($P_6$,paw)-free graphs.
\end{theo}

{\bf Proof.} Let $G$ be a graph of $n$ vertices. Let {\bf y} be any rational $n$-vector. Let us prove that when $G$ is ($P_6$,triangle)-free and more in general ($P_6$,paw)-free there exists a polynomial time algorithm that either asserts that {\bf y} belongs to STAB($G$) or finds a facet-defining inequality of STAB($G$) violated by {\bf y}.

First assume that $G$ is ($P_6$,triangle)-free. The nonnegativity constraints can checked by substitution. So one may assume that the entries of {\bf y} are nonnegative. By Theorem \ref{theo final final}, the remaining facet-defining inequalities of STAB($G$) are given by the set $\{\Phi (H) : H \in \{K_2,C_5,G_1,\ldots,G_{24}\}\}$. Then, since the cardinality of $\{\Phi (H) : H \in \{K_2,C_5,G_1,\ldots,G_{24}\}\}$ is bounded by a constant, such inequalities can be checked by substitution.

Then assume that $G$ is ($P_6$,paw)-free. The nonnegativity constraints can be checked by substitution. So one may assume that the entries of {\bf y} are nonnegative. By Corollary \ref{coro final final}, the remaining facet-defining inequalities of STAB($G$) are given by the set $\{\Phi (H) : H \in C^*(K_2,G) \cup C^*(C_5,G) \cup C^*(G_1,G) \cup \ldots \cup C^*(G_{24},G)\}$. The inequalities from $C^*(K_2,G)$ can be checked by solving the maximum weight clique problem in $G$ (weighted by {\bf y}): as remarked above, this can be done in polynomial time. Then assume that {\bf y} satisfies also the inequalities from $C^*(K_2,G)$. Then every (not necessarily maximal) clique inequality is satisfied.
This implies that, if {\bf y} violates an inequality over the remaining facet-defining inequalities of STAB($G$), then there exists a facet-defining inequality of STAB($G[H]$) for $H \in T = \{C_5,G_1,\ldots,G_{24}\}$ which is violated as well (that comes also from the structure of the elements of $C^*(H,G)$). As shown in Theorem \ref{theo final}, the facet-inducing subgraphs of a graph $H \in T$ are contained in $T \cup \{K_2\}$. Then the set of facet-defining inequalities of STAB($G[H]$) for $H \in T$ is contained in $\{\Phi (H) : H \in \{K_2,C_5,G_1,\ldots,G_{24}\}\}$. Then, since the cardinality of $\{\Phi (H) : H \in \{K_2,C_5,G_1,\ldots,G_{24}\}\}$ is bounded by a constant, such inequalities can be checked by substitution.  \qed

\section{A note on new facet-inducing graphs}

In this section let point out some peculiarities of new facet-inducing graphs detected along this study with the help of a software according to Remark 3.
Graphs $H_1,H_2,H_3$, drawn respectively in Figure 2, 3, 4, are (prime) facet-inducing. In particular let us list some peculiarities of graphs $H_1$ and $H_2$.   \\

{\bf Graph $H_1$.}
\begin{itemize}
\item[$(i)$] $H_1$ is facet-inducing;
\item[$(ii)$] $H_1 - \{v\}$ is facet-inducing for every vertex $v$ of $H_1$;
\item[$(iii)$] STAB($H_1 - \{16\}$) has 641 full facets, i.e., $|\Phi (H_1 - \{16\})| = 641$.
%\item[$(iv)$] $H_1$ has several facet-inducing subgraphs: for example, $H_1 - \{1,2,3,4,5\}$ and $H_1 - \{1,2,3,4,5,16\}$ which have respectively a unique full facet, i.e., a rank facet of right hand side 3.
\end{itemize}

{\bf Graph $H_2$.}
\begin{itemize}
\item[$(iv)$] $H_2$ is facet-inducing;
\item[$(v)$] STAB($H_2$) has 26617 facets;
%this seems to be linked to the "symmetric" structure of $H_2$; $H_2$ has several facet-inducing subgraphs $F$, i.e., graphs $K_2,C_5,G_1,\ldots,G_{13}$, with $|\Phi (F)|$ often greater than 10 (appearing a large number of times in $H_2$);
\item[$(vi)$] $H_2 - \{u\}$ is isomorphic to $H_2 - \{v\}$ for every pair of vertices $u,v$ of $H_2$;
\item[$(vii)$] $H_2 - \{u,v\}$ is isomorphic to $H_2 - \{u',v'\}$ for every pair of disjoint edges $uv$, $u'v'$ of $H_2$;
\item[$(viii)$] $H_2 - \{v\}$ is facet-inducing for every vertex $v$ of $H_2$;
\item[$(ix)$] $H_2 - \{u,v\}$ is facet-inducing for every edge $uv$ of $H_2$;
\item[$(x)$] $H_2$ seems to enjoy further properties, such as regularity ...
\end{itemize}

Concerning ($ii$), ($viii$) and ($ix$), in a not expert and poor knowledge we ignore any other facet-inducing graph with those properties (apart from cliques); concerning ($iii$), similarly we ignore any other facet-inducing graph, of that low order, whose stable set polytope has a so large number of full facets; concerning ($v$), similarly we ignore any other facet-inducing graph, of that low order, whose stable set polytope has a so large number of facets.

%The third author in his poor knowledge ignores: concerning ($ii$), ($viii$) and ($ix$), any other facet-inducing graph with those properties (apart from cliques); concerning ($iii$), any other facet-inducing graph, of that low order, whose stable set polytope has a so large number of full facets; concerning ($v$), any other facet-inducing graph, of that low order, whose stable set polytope has a so large number of facets.

Graphs $H_1$ and $H_2$ necessarily have a symmetric structure. In particular maybe one could expect that graphs $H_1$ and $H_2$ could lead to new families of (prime) facet-inducing graphs. To this end let us try to point out a symmetric representation for graphs $H_1$ and $H_2$. Concerning graph $H_1$, maybe Figure 2 seems to be enough in this sense since it shows a symmetry based on three $C_5$'s (one of which with a bigger out-degree) plus one external vertex, namely vertex 16. Concerning graph $H_2$, also recalling that $H_2$ is regular, let us try to provide below a highly symmetric representation. Let us refer to Figure 6.

\begin{figure}
\centering
\includegraphics[width=\textwidth]{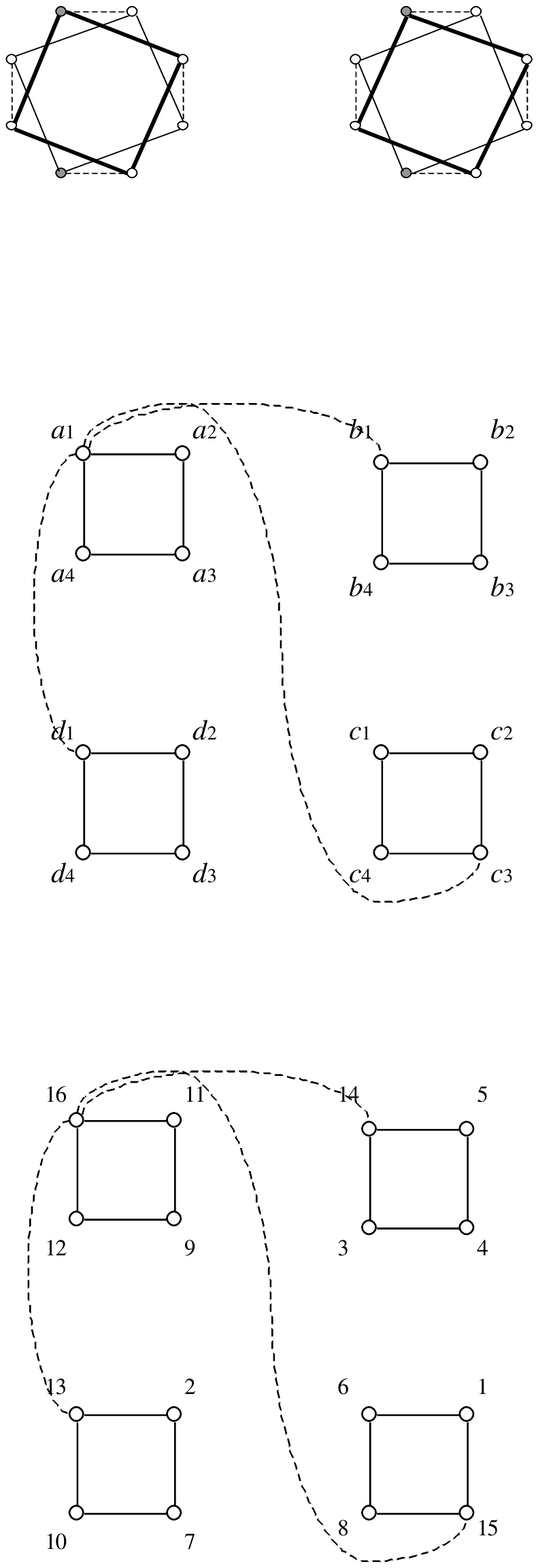}
\caption{Representations of graph $H_2$}
\end{figure}

As a preliminary let us observe that, it seems that there is no ordering of vertices of graph $H_2$, say $v_1,\ldots,v_{16}$, such that each vertex $v_i$ admits a list of neighborhoods formed by those vertices $v_{i+j}$ which are at the same fixed distances (sum taken modulo 16).

A first representation of graph $H_2$ may be obtained by considering two copies, say $A$ and $B$, of a graph of 8 vertices (Figure 6 up). The vertex-set of $A$ (of $B$) can be uniquely partitioned into pairs, formed by vertices which have the maximum distance each other in $A$ (in $B$). One of such a pair in $A$ (in $B$) is distinguished in Figure 6 up by color gray. Then add edges in order to have a join from respectively each such a pair in $A$ to its homologous pair in $B$.

A second representation of graph $H_2$ may be obtained by considering a $C_4$, say $C$, of vertices $a,b,c,d$ and of edges $ab,bc,cd,da$.
Then expand each vertex of $C$ into a $C_4$, that is for $x = a,b,c,d$, expand vertex $x$ into a $C_4$ of vertices $x_1,x_2,x_3,x_4$ and edges $x_1x_2,x_2x_3,x_3x_4,x_4x_1$.
Then add the following edges: for $i = 1,2,3,4$, add edges $a_ib_i,b_ic_i,c_id_i,d_ia_i$, i.e., add edges between homologous vertices of the $C_4$'s when such $C_4$'s correspond to adjacent vertices of $C$, and add edges $a_ic_{i+2},b_id_{i+2}$ (sum taken modulo 4), i.e., add edges between oppositive vertices of the $C_4$'s when such $C_4$'s correspond to nonadjacent vertices of $C$ (Figure 6 middle). In particular the isomorphism from this representation to graph $H_2$ of Figure 6 is based on the following bijection between vertex-sets: from $a_1,a_2,a_3,a_4,b_1,b_2,b_3,b_4,c_1,c_2,c_3,c_4,d_1,d_2,d_3,d_4$ to $16,11,9,12,14,5,4,3,6,1,15,8,13,2,7,10$ (Figure 6 down). \\

{\bf Acknowledgement.} I would like to thank Prof. Gianluca Amato, Prof. Marco Dall'Aglio, and Prof. Caterina De Simone for the help given to use the software Mathematica and the software mentioned in Remark 3. Then would like to thank Prof. Caterina De Simone for having improved the presentation of the proof of Proposition 1 of \cite{Mosca2008}, i.e., of Proposition 2. 

Then would like to thank for different reasons many persons which I met in these years of study, in Italy, Prof. Claudio Arbib for having introduced me to this study, for his support, and for having taught me many things, Prof. Giandomenico Boffi, Prof. Marco Dall'Aglio, Prof. Caterina De Simone, Prof. Paolo Nobili, Prof. Antonio Sassano, out of Italy, Prof. Andreas Brandst\"adt for many reasons detailed below, Prof. Ho\`ang-Oanh Le, Prof. Van Bang Le, Prof. Vadim V. Lozin for many cooperations so instructive to me, Prof. Haiko M\"uller, and generally other persons.

Then would like to thank in a special way Prof. Andreas Brandst\"adt for his support at a moment in which I had no position and no support, for having taught me many things, for many cooperations so instructive to me, and for having been a lighten benchmark under different aspects. 
 
Finally would like humbly to dedicate this paper to (the memory of) my parents, namely, Giuseppe Mosca and Meris Ciufolini.

%in particular, Prof. Andreas Brandst\"adt and Prof. Vadim V. Lozin, for many cooperations very enriching to me.

Just try to pray a lot and am able to do nothing without that. 

%Salmo 118,1

%Ad laudem Domini.

%{\bf Acknowledgement 2.} Let me testify that I try to pray a lot and am able to do nothing without that.

%{\bf Acknowledgement 3.} Ad Laudem Domini.

%{\bf Acknowledgement 5.} Finally would like to witness that in my nullity I try to pry a lot and am able to do nothing without that.

\begin{footnotesize}
\renewcommand{\baselinestretch}{0.4}

\end{footnotesize}

\end{document}